\documentclass[12pt, onecolumn]{IEEEtran}

\usepackage{graphics,graphicx} % for pdf, bitmapped graphics files
\usepackage{epsfig} % for postscript graphics files

\usepackage{amsmath}
\usepackage{amssymb}
\usepackage{tikz}
\usepackage{amsthm}
\usepackage{comment}
\usepackage[export]{adjustbox}
\usepackage{mathcomSTEP}
\usepackage{subcaption}
\usepackage{setspace}	
\usepackage[utf8]{inputenc}
\usepackage{cite}
\usepackage{float}
\usepackage{lipsum}
\usepackage{stfloats}

\usetikzlibrary{positioning}
\allowdisplaybreaks

\newtheorem{prop}{Proposition}

\theoremstyle{remark}

\newtheorem{definition}{Definition}

\newcommand{\Wt}{\widetilde{W}}
\newcommand{\Nrx}{N_{\rm r} }
\newcommand{\Ntx}{N_{\rm t} }

\usepackage{tikz}
\usetikzlibrary{positioning}
\usetikzlibrary{arrows,fit}
\usetikzlibrary{shapes.geometric}
\usetikzlibrary{positioning,calc}
\usetikzlibrary{decorations.markings,shapes,arrows}

\usepackage{epstopdf}
\usepackage{pgfplots,setspace}
\usepackage{circuitikz}

\tikzstyle{int}=[draw, fill=blue!10, minimum height = .5 cm, minimum width=1 cm,thick ]
\tikzstyle{int1}=[draw,  minimum height = .15 cm, minimum width=1 cm,thick ]
\tikzstyle{sum}=[circle, fill=blue!10, draw=black ]

\allowdisplaybreaks

\title{
A General Framework for \\
MIMO Receivers with Low-Resolution Quantization
}
\author{

	\IEEEauthorblockN{Stefano Rini\IEEEauthorrefmark{1}, Luca Barletta\IEEEauthorrefmark{2}, Yonina C. Eldar\IEEEauthorrefmark{3}, and Elza Erkip\IEEEauthorrefmark{4}\\}
	\IEEEauthorblockA{
        \IEEEauthorrefmark{1}
		National Chiao Tung University, Hsinchu, Taiwan\\
	}
	\and
	\IEEEauthorblockA{
        \IEEEauthorrefmark{2}
		Politecnico di Milano, Milano, Italy\\
	}
	\and
	\IEEEauthorblockA{
\IEEEauthorrefmark{3}
		Technion, Haifa, Israel\\
	}
	\and
	\IEEEauthorblockA{
\IEEEauthorrefmark{4}
		NYU Tandon School of Engineering, New York, USA
	}
	
}

\begin{document}

\maketitle
\begin{abstract}
The capacity of a discrete-time multi-input multi-output (MIMO) Gaussian channel with output quantization is investigated for different
 receiver architectures.
A general formulation of this problem is proposed in which the antenna outputs are processed by analog combiners while
sign quantizers are used for analog-to-digital conversion.
%
%The configuration of the analog combiners is chosen as a function of the channel realization, so that
% capacity can be maximized over the set of available configurations for the given number of sign quantizers.
%
To exemplify this approach, four analog receiver architectures of varying generality and complexity are considered:
(a) multiple antenna selection and sign quantization of the antenna outputs,
(b) single antenna selection and multilevel quantization,
(c) multiple antenna selection and multilevel quantization, and
(d) linear combining of the antenna outputs and multilevel quantization.
Achievable rates are studied as a function of the number of
available sign quantizers and compared among different architectures.
In particular, it is shown that architecture (a) is sufficient to attain the optimal high signal-to-noise ratio performance for  a MIMO receiver in which the number of antennas is larger than the number of sign quantizers.
Numerical evaluations of the average performance are presented for the case in which the channel gains
are i.i.d. Gaussian.
%
%By comparing the largest attainable rates for a given number of sign quantizers,
%it is possible to quantify the limiting rate advantages provided by each of the receiver analog architectures.
%%
%%In particular, it is shown that sign quantization of the antenna outputs is sufficient, among all possible receiver architectures studied, to attain the optimal high signal-to-noise ratio (SNR) performance for  MIMO receiver in which the number of antennas is larger than the number of sign quantizers.
%\blue{SR: complete}
\end{abstract}

\section{Introduction}
Low-resolution quantization is an important technology for massive MIMO and millimeter-wave communication systems as it allows the transceivers to operate at low power levels
\cite{swindlehurst2014millimeter}.

%
%The coupling of  low-power, low-complexity and low-resolution quantizers with large antenna arrays holds the promise of a multi-fold increase in  network throughput \cite{swindlehurst2014millimeter}.
%
Although the performance of MIMO receivers with large antenna arrays and low-resolution quantizers has been investigated in the literature
under different assumptions on the hardware limitations and antenna architectures,
% \cite{alkhateeb2014mimo},
%
%Unfortunately though,
a complete fundamental information theoretic understanding
%of  hybrid analog/digital receiver architectures
is currently not available.
In this paper, we propose a unified framework to analyze and compare low-resolution receiver architectures.
%under various constraints on the
% analog combining capabilities.
%
More specifically, we assume that the receiver is comprised of  $N_{SQ}$ sign quantizers that process $\Nrx$ antenna outputs.
Each sign quantizer is connected to the antenna outputs via an analog combining circuit with limited processing capabilities.
Through this general formulation, we study
the effects of limited processing and low-resolution quantization
on the capacity of MIMO channels.
%Given a configuration of the analog combiners,  the sign quantizers produce a binary vector of length $N_{SQ}$.
%%
%The configuration of the analog combiners is allowed to depend on the channel realization, so that
% capacity can be maximized  over the possible configurations
%%
%subject to the assumed hardware constraints.
%
%The underlying assumptions in our problem formulation are as follows:
Op-amp voltage comparators are employed
in nearly all analog-to-digital converters to obtain multilevel
quantization. Given the receiver's ability to partially reconfigure its circuitry depending on the channel realization, it is of interest to determine which
configuration of the comparators  yields the
largest capacity.

\subsubsection*{Literature Review}
%
%Many authors have investigated  the effects of quantization on MIMO systems:
Quantization in MIMO systems is a well-investigated topic in the literature:
for the sake of brevity we focus here on the results
regarding sign quantization.\footnote{
In the literature, the term ``one-bit quantization'' most often refers to sign quantization of the antenna outputs.
Here, as in \cite{koch2013low},  we prefer the term ``sign quantization''  since we distinguish between sign and threshold quantization.}
The authors in~\cite{nossek2006capacity} are perhaps the first to point out that the capacity loss in MIMO channels due to coarse quantization is surprisingly small, although this observation is supported mostly through numerical evaluations.
%
%Binary  space-time codes for MIMO systems with quantized outputs are considered in \cite{ivrlac2006challenges}: the authors show that almost all codes behave poorly under coarse quantization with only a few exceptions.
%%
%In \cite{mezghani2007ultra},  the authors study the second-order expansion
%of the MIMO channel capacity with sign quantization and determine the low SNR performance.
%
In \cite{singh2009limits}, the authors derive fundamental properties of the capacity-achieving
distribution for a single-input single-output (SISO) channel with output quantization.
A lower bound on the capacity of sign-quantized MIMO channels with Gaussian inputs
based on the Bussgang decomposition is derived in \cite{mezghani2012capacity}.
The high signal-to-noise ratio (SNR) asymptotics for complex MIMO channels with sign quantization are studied are \cite{mo2014high}.
For the SISO channel with  threshold quantization, \cite{koch2013low} shows that, in the limit of vanishing SNR,
asymmetric quantizers outperform symmetric ones.
%
%In the presence of fading, low-precision quantization affects the receiver's performance by also hampering channel estimation,
%as investigated in \cite{mezghani2008analysis,krone2010fading,risi2014massive}.
%
%Although
%
%The  fading  MIMO channel  with sign quantization of the channel output was first considered in \cite{mezghani2008analysis}
%for the case of quasi-static Rayleigh fading.
%% and derive the capacity for the SISO case.
%%
%Optimal modulation schemes, ergodic capacity and the outage probability for this model are further investigated in \cite{krone2010fading}.
%%
%In \cite{risi2014massive}, a transmission scheme
%%involving
%%s \markL{LUCA: maybe you missed a word here}
% based on maximal ratio combining, zero-forcing and least squares detection is derived.
%

\subsubsection*{Contributions}
We focus, in the following, on four analog receiver architectures with different levels of complexity: (a)~multiple antenna selection and sign quantization, (b)~single antenna selection and multilevel quantization, (c)~multiple antenna selection and multilevel quantization, and (d)~linear combining and multilevel quantization.
The architecture (c) is more general than both (a) and (b), and (d) is the most general one.
We study the case of a SIMO channel and a MIMO channel and provide capacity bounds of each architecture as a function of the
% the optimal performance as a function of the
%average transmit power, channel realization, and
number of sign quantizers.
%
%Investigating the SIMO case separately allows us to get tighter results.
%
For the SIMO channel, our results suggest conditions under which the capacity of the architecture with multiple antenna selection and multilevel quantization
closely approaches that of the architecture with linear combining and multilevel quantization.
For the MIMO channel with linear combining and multilevel quantization, we derive an approximatively optimal usage of the sign quantizers
as a variation of the classic water-filling power allocation scheme.
%for a given channel realization and average transmit power.
%
This solution shows that, if the number of antennas at the receiver is larger than the number of sign quantizers,
sign quantization is sufficient to attain the optimal performance in the high SNR regime.
%
%This result partially motivates the study of sign quantization for MIMO systems in the high SNR regime as in~\cite{mo2014high}.
%
%The average performance of the different architectures is numerically evaluated for the case in which the channel gains
Numerical evaluations are provided for the case in which the channel gains are i.i.d. Gaussian distributed.
\subsubsection*{Paper Organization}
%
%The remainder of the paper is organized as follows:
%
Sec.~\ref{sec:Channel Model} introduces the channel model.
Sec.~\ref{sec:Known results} reviews the results available for the case of sign quantization of the channel outputs.
%%%
%%An example motivating the problem formulation is presented in Sec. \ref{sec:Motivating Example}.
%%%
The main results  are given in Sec.~\ref{sec:Main Results}.
Numerical evaluations are provided in Sec.~\ref{sec:Numerical Evaluations}.
Sec.~\ref{sec:Conclusion} concludes the paper.

\subsubsection*{Notation}
We adopt the standard notation for $H_2(x)=-x\log x-(1-x)\log(1-x)$ and $Q(x)=1/\sqrt{2 \pi} \int_x^{+\infty} \exp(-u^2/2)\diff u$.
All logarithms are taken in base two.
For the SISO model, we set  $\Hv=1$ w.l.o.g.,  for the MISO and SIMO models we denote the channel matrix
as $\hv$ and $\hv^T$ respectively.
%For these models, we also assume  $h_1 \geq h_2 \geq h_3 \ldots$ w.l.g.
%
For the MIMO case, the vector $\lav=[\la_1 \ldots \la_{\min\{\Ntx,\Nrx\}}]$ contains the eigenvalues of the matrix $\Hv\Hv^T$.
The  identity matrix of size $n\times n$  is indicated as $\Iv_{n}$, the  all-zero/all-one matrix of size $n \times m$
as $\zerov_{n \times m}$/$\onev_{n \times m}$.
Finally,  $\Pcal_{\pi}$ indicates the set of all permutation matrices.

\begin{figure*}
\vspace{-1cm}
\captionsetup[subfigure]{justification=centering}
 \begin{subfigure}[c]{0.3 \textwidth}
        \resizebox{5.5 cm }{!}{
\begin{tikzpicture}[node distance=2.5cm,auto,>=latex]
\node at (0,2) {};
\node  at (0.25,3) (y1) {$W_1$};
\node  at (0.25,2) (y2) {$W_2$};
\node  at (0.25,1) (y3) {$W_3$};
\node  at (0.25,0) (y4) {$W_4$};
\node  at (1,3) (d1) [sum,scale=0.5]{};
\node  at (1,2) (d2) [sum,scale=0.5]{};
\node  at (1,1) (d3) [sum,scale=0.5]{};
\node  at (1,0) (d4) [sum,scale=0.5]{};
\draw (y1) [line width=1 pt] -> (d1);
\draw (y2) [line width=1 pt] -> (d2);
\draw (y3) [line width=1 pt] -> (d3);
\draw (y4) [line width=1 pt] -> (d4);
\node  at (2,3) (e1) [sum,scale=0.5]{};
\node  at (2,2) (e2) [sum,scale=0.5]{};
\node  at (2,1) (e3) [sum,scale=0.5]{};
\draw [<->] (2,3+0.25) to[out=180,in=120, distance=.25cm ] (1.75,3-0.25);
\draw [<->] (2,2+0.25) to[out=180,in=120, distance=.25cm ] (1.75,2-0.25);
\draw [<->] (2,1+0.25) to[out=180,in=120, distance=.25cm ] (1.75,1-0.25);
\node  at (3.5,3) (quant1) [int1] {$\sign(\cdot)$};
\node  at (3.5,2) (quant2) [int1] {$\sign(\cdot)$};
\node  at (3.5,1) (quant3) [int1] {$\sign(\cdot)$};

\node  at (5,3) (t1) {$Y_1$};
\node  at (5,2) (t2) {$Y_2$};
\node  at (5,1) (t3) {$Y_3$};
\draw (quant1) [->,line width=1 pt] -> (t1) ;
\draw (quant2) [->,line width=1 pt] -> (t2);
\draw (quant3) [->,line width=1 pt] -> (t3);
\draw (e1) [->,line width=1 pt] -> (quant1) ;
\draw (e2) [->,line width=1 pt] -> (quant2);
\draw (e3) [->,line width=1 pt] -> (quant3);
\draw (d1) [line width=1 pt] -> (e1);
\draw (d3) [line width=1 pt] -> (e2);
\draw (d4) [line width=1 pt] -> (e3);
\node at (0,-1) {};
 \end{tikzpicture}
}
\vspace{-1.5cm}
\caption{Multiple antenna selection and sign quantization.}
\label{fig:channel model sign}
 \end{subfigure}
\begin{subfigure}[c]{0.373 \textwidth}
        \resizebox{6.3 cm }{!}{
\begin{tikzpicture}[node distance=2.5cm,auto,>=latex]
\tikzset{point/.style={coordinate},
block/.style ={draw, thick, rectangle, minimum height=4em, minimum width=6em},
line/.style ={draw, very thick,-},
}

\node  at (0.25,3) (y1) {$W_1$};
\node  at (0.25,2) (y2) {$W_2$};
\node  at (0.25,1) (y3) {$W_3$};
\node  at (0.25,0) (y4) {$W_4$};
\node  at (1,3) (d1) [sum,scale=0.5]{};
\node  at (1,2) (d2) [sum,scale=0.5]{};
\node  at (1,1) (d3) [sum,scale=0.5]{};
\node  at (1,0) (d4) [sum,scale=0.5]{};
\draw (y1) [line width=1 pt] -> (d1);
\draw (y2) [line width=1 pt] -> (d2);
\draw (y3) [line width=1 pt] -> (d3);
\draw (y4) [line width=1 pt] -> (d4);
\node  at (2,3) (e1) [sum,scale=0.5]{};
\node  at (2,2) (e2) [sum,scale=0.5]{};
\node  at (2,1) (e3) [sum,scale=0.5]{};
\draw [<->] (2,3+0.25) to[out=180,in=120, distance=.25cm ] (1.75,3-0.25);
\draw [<->] (2,2+0.25) to[out=180,in=120, distance=.25cm ] (1.75,2-0.25);
\draw [<->] (2,1+0.25) to[out=180,in=120, distance=.25cm ] (1.75,1-0.25);

\node  at (6,3) (t1) {$Y_1$};
\node  at (6,2) (t2) {$Y_2$};
\node  at (6,1) (t3) {$Y_3$};
\node  at (3,3) (e11) [sum,scale=0.5]{$+$};
\node  at (3,2) (e21) [sum,scale=0.5]{$+$};
\node  at (3,1) (e31) [sum,scale=0.5]{$+$};
\node  at (4.5,3) (quant1) [int1] {$\sign(\cdot)$};
\node  at (4.5,2) (quant2) [int1] {$\sign(\cdot)$};
\node  at (4.5,1) (quant3) [int1] {$\sign(\cdot)$};
\draw (e11) [line width=1 pt] -> (quant1);
\draw (e21) [line width=1 pt] -> (quant2);
\draw (e31) [line width=1 pt] -> (quant3);
\draw (quant1) [->,line width=1 pt] -> (t1) ;
\draw (quant2) [->,line width=1 pt] -> (t2);
\draw (quant3) [->,line width=1 pt] -> (t3);
\draw (y1) [line width=1 pt] -> (d1);
\draw (y2) [line width=1 pt] -> (d2);
\node  at (2.5,3-.5) (t11) {$t_1$};
\node  at (2.5,2-.5) (t21) {$t_2$};
\node  at (2.5,1-.5) (t31) {$t_3$};
\node  at (2.5,2) (t111) {};
\node  at (2.5,-1) (t211) {};
\draw (t11)[->,line width=1 pt] -| (e11)   ;
\draw (t21) [->,line width=1 pt] -| (e21) ;
\draw (t31) [->,line width=1 pt] -| (e31) ;
\draw (e1) [->,line width=1 pt] -> (e11);
\draw (e2) [->,line width=1 pt] -> (e21);
\draw (e3) [->,line width=1 pt] -> (e31);
\draw (d2) [line width=1 pt] -> (e1);
\draw (d2) [line width=1 pt] -> (e2);
\draw (d2) [line width=1 pt] -> (e3);
)
 \end{tikzpicture}
 }
 \vspace{-1cm}
 \caption{Single antenna selection and\\ multilevel quantization.}
 \label{fig:single antenna}
  \end{subfigure}
  \begin{subfigure}[c]{0.373 \textwidth}
        \resizebox{6.3 cm }{!}{
\begin{tikzpicture}[node distance=2.5cm,auto,>=latex]
\tikzset{point/.style={coordinate},
block/.style ={draw, thick, rectangle, minimum height=4em, minimum width=6em},
line/.style ={draw, very thick,-},
}

\node  at (0.25,3) (y1) {$W_1$};
\node  at (0.25,2) (y2) {$W_2$};
\node  at (0.25,1) (y3) {$W_3$};
\node  at (0.25,0) (y4) {$W_4$};
\node  at (1,3) (d1) [sum,scale=0.5]{};
\node  at (1,2) (d2) [sum,scale=0.5]{};
\node  at (1,1) (d3) [sum,scale=0.5]{};
\node  at (1,0) (d4) [sum,scale=0.5]{};
\draw (y1) [line width=1 pt] -> (d1);
\draw (y2) [line width=1 pt] -> (d2);
\draw (y3) [line width=1 pt] -> (d3);
\draw (y4) [line width=1 pt] -> (d4);
\node  at (2,3) (e1) [sum,scale=0.5]{};
\node  at (2,2) (e2) [sum,scale=0.5]{};
\node  at (2,1) (e3) [sum,scale=0.5]{};
\draw [<->] (2,3+0.25) to[out=180,in=120, distance=.25cm ] (1.75,3-0.25);
\draw [<->] (2,2+0.25) to[out=180,in=120, distance=.25cm ] (1.75,2-0.25);
\draw [<->] (2,1+0.25) to[out=180,in=120, distance=.25cm ] (1.75,1-0.25);

\node  at (6,3) (t1) {$Y_1$};
\node  at (6,2) (t2) {$Y_2$};
\node  at (6,1) (t3) {$Y_3$};
\node  at (3,3) (e11) [sum,scale=0.5]{$+$};
\node  at (3,2) (e21) [sum,scale=0.5]{$+$};
\node  at (3,1) (e31) [sum,scale=0.5]{$+$};
\node  at (4.5,3) (quant1) [int1] {$\sign(\cdot)$};
\node  at (4.5,2) (quant2) [int1] {$\sign(\cdot)$};
\node  at (4.5,1) (quant3) [int1] {$\sign(\cdot)$};
\draw (e11) [line width=1 pt] -> (quant1);
\draw (e21) [line width=1 pt] -> (quant2);
\draw (e31) [line width=1 pt] -> (quant3);
\draw (quant1) [->,line width=1 pt] -> (t1) ;
\draw (quant2) [->,line width=1 pt] -> (t2);
\draw (quant3) [->,line width=1 pt] -> (t3);
\draw (y1) [line width=1 pt] -> (d1);
\draw (y2) [line width=1 pt] -> (d2);
\node  at (2.5,3-.5) (t11) {$t_1$};
\node  at (2.5,2-.5) (t21) {$t_2$};
\node  at (2.5,1-.5) (t31) {$t_3$};
\node  at (2.5,2) (t111) {};
\node  at (2.5,-1) (t211) {};
\draw (t11)[->,line width=1 pt] -| (e11)   ;
\draw (t21) [->,line width=1 pt] -| (e21) ;
\draw (t31) [->,line width=1 pt] -| (e31) ;
\draw (e1) [->,line width=1 pt] -> (e11);
\draw (e2) [->,line width=1 pt] -> (e21);
\draw (e3) [->,line width=1 pt] -> (e31);
\draw (d1) [line width=1 pt] -> (e1);
\draw (d2) [line width=1 pt] -> (e2);
\draw (d2) [line width=1 pt] -> (e3);
)
 \end{tikzpicture}
 }
 \vspace{-1cm}
 \caption{Multiple antenna selection and\\ multilevel quantization.}
 \label{fig:mulitiple antenna}
  \end{subfigure}
\caption{
Different  analog receiver architectures.
}
\label{fig:channel model}
\vspace{-.5 cm }
\end{figure*}
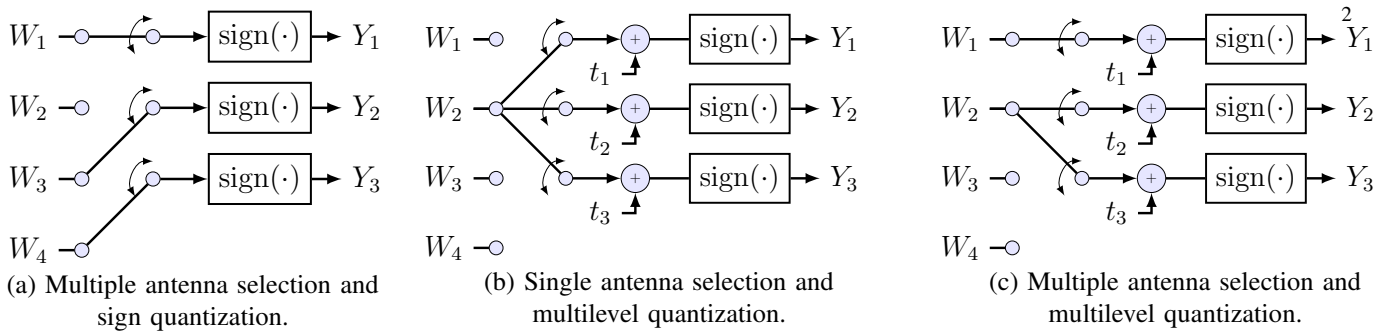

\section{Channel Model}
\label{sec:Channel Model}
\subsubsection*{Problem Formulation}
\label{sec:Problem Formulation}

We consider a discrete-time real-valued MIMO channel with $\Ntx$ transmit antennas and $\Nrx$ receive antennas.
At the $n^{\rm th}$ channel use, the antenna output vector $\Wv_n=[W_{1,n} \ldots W_{\Nrx,n}]^T$,
is obtained from the channel input vector $\Xv_n=[X_{1,n} \ldots X_{\Ntx,n}]^T$ as
%channel output vector $\Wv$ is obtained from the channel input vector $\Xv$ as
\ea{
\Wv_n=\Hv \Xv_n+\Zv_n, \quad  n\in [1 \ldots N],
\label{eq:antenna out}
}
where $\Hv$ is a full rank matrix of size $\Nrx \times \Ntx$
% with linearly independent rows,
\footnote{This condition guarantees the existence of a right pseudo-inverse for $\Hv$ and holds with high probability in a richly scattering environment.}
%, known at both transmitter and receiver,
 and $\Zv_n$ is an $\Nrx$-vector of i.i.d. additive Gaussian noise samples with zero mean and unitary variance.
 The channel matrix $\Hv$ is assumed to be known at both transmitter and receiver and to be fixed throughout the transmission block-length $N$.
 %i.i.d. Gaussian variables of zero mean and unit variance.
%
%The channel matrix is assumed full-rank
%For a given transmission block-length,
The channel input vector is subject to the average power constraint $\sum_{n=1}^N \Ebb[ |\Xv_{n}|_2^2]\leq  N P$ where $|\Xv_{n}|_2$ indicates the 2-norm.

The antenna output vector is processed through $N_{SQ}$ sign quantizers, each receiving a linear combination of the antenna output vector plus a constant,\footnote{It must be noted that generating a precise voltage reference is another major hurdle in analog-to-digital conversion. Although possible in our framework, in the following we do not consider such limitation.} \emph{i.e.}
\ea{
\Yv_n = \sign(\Vv \Wv_n+\tv), \quad  n\in [1 \ldots N],
\label{eq:quantized model}
}
where $\Vv$ is the analog combining matrix of size $N_{SQ} \times \Nrx$, $\tv$~is a threshold vector of length $N_{SQ}$ and
 $\sign(\uv)$ is the function producing the sign of each component of the vector $\uv$ as plus or minus one, so that $\Yv_n \in \{-1,+1\}^{N_{SQ}}$.
For a given choice of combining matrix $\Vv$ and threshold vector $\tv$, the capacity of the model in \eqref{eq:quantized model} is given by
\ea{
\Ccal(\Vv,\tv)=\max_{P_{\Xv}(\xv), \ \Ebb[ |\Xv|_2^2]\leq  P} I(\Xv;\Yv),
\label{eq:capacity V t}
}
where we have explicitly expressed the dependency of the capacity on the parameters $\{\Vv,\tv\}$.\footnote{The capacity $\Ccal(\Vv,\tv)$
is also a function of the channel matrix $\Hv$, although not explicitly indicated.
 %which is assumed to be known at both the transmitter and the receiver.
 }
The analog processing capabilities at the receiver are modeled as a set of feasible values of $\{\Vv,\tv\}$, denoted as  $\Fcal$.
%
%By maximizing the capacity expression in \eqref{eq:capacity V t} over $\Fcal$, we obtain the maximum capacity over $\Fcal$, that is
Our goal is to maximize the capacity expression in \eqref{eq:capacity V t} over $\Fcal$, namely
%, ie
\ea{
\Ccal(\Fcal)=\max_{\{\Vv,\tv\} \in \Fcal} \Ccal(\Vv,\tv).
\label{eq:max c}
}

\subsubsection*{Relevant Architectures}
\label{sec:Relevant Architectures}

The formulation in \eqref{eq:max c} attempts to capture the tension  between
the quantization of few antennas with high precision versus the quantization of many antennas with low precision.
This is accomplished by treating the sign quantizers as a resource to be allocated optimally among a set of possible configurations $\cal{F}$.
Note that  \mbox{$M$-level} multilevel quantization  can be obtained by using $M-1$ sign quantizers and appropriate thresholds $\tv$, resulting in $\log(M)$ information bits.
%
%the quantizer output value can be represented using $\log(M)$ information bits (all logarithms are taken in base two).
%
It follows that sign quantization produces the most information bits per sign quantizer and increasing the
number of quantization levels increases the information bits only logarithmically.

To exemplify the insights provided by our approach,
%To exemplify the proposed approach, we focus on four
%To further investigate the role finite precision quantization on the MIMO capacity,
we study four analog receiver architectures:
%

%\smallskip
\noindent
{\bf (a) Multiple antenna selection and sign quantization:}
Here %each quantizer produces the sign of an antenna output, so that
 $\Fcal$ in \eqref{eq:max c} is selected as
\ea{
\Fcal_{a}& =\lcb \Vv=\lsb \Iv_{N_{SQ}},\zerov_{N_{SQ} \times (\Nrx-N_{SQ})} \rsb P_{\pi}, \ P_{\pi} \in \Pcal_{\pi}, \rnone \nonumber \\
  & \quad \ \  \lnone \ \tv=\zerov_{N_{SQ}\times 1} \rcb,
\label{eq:sign quantization}
 }
that is, each sign quantizer is connected to one of the channel outputs.
%, both of dimensions specified in the subscript.
Figure \ref{fig:channel model sign} represents this model for $\Nrx=4$ and $N_{SQ}=3$.
%

%
%\smallskip
\noindent
{\bf (b) Single antenna selection and multilevel quantization:}
%
%here all the quantizers are connected to the same antenna output, providing an $(N_{SQ}+1)$-level
%quantization of the output.
%
For this receiver architecture, the sign quantizers are used to construct an $(N_{SQ}+1)$-level quantizer:
\ea{
\Fcal_{b}& =
\lcb
 \Vv  = \lsb \onev_{N_{SQ}  \times  1 }, \zeros_{ N_{SQ} \times (\Nrx-1)} \rsb P_{\pi}, \ P_{\pi} \in \Pcal_{\pi}, \rnone \nonumber \\
    & \quad \quad \  \lnone \tv \in \Rbb^{N_{SQ}} \rcb,
 \label{eq:th quant}
%\}.
}
Figure \ref{fig:single antenna} shows this model for $\Nrx=4$ and $N_{SQ}=3$.

%\smallskip
\noindent
{\bf (c) Multiple antenna selection and multilevel quantization:}
Here, each sign quantizer can select an antenna output and a voltage offset before performing quantization.
%
%For this receiver architecture, we have
This is obtained by choosing
\ea{
\Fcal_{c} & =
\lcb \Vv \ST V_{ij} \in \{0,1\}, \ \dsum_{j=1}^{\Nrx} V_{ij}=1, \
% \rnone \nonumber \\
    %& \quad \quad  \ \lnone
    \tv \in \Rbb^{N_{SQ}} \rcb.
 \label{eq:th quant}
}
%\markL{LUCA: I think that the pre-multiplication by $Q_\pi$ has no effect, because all rows of $\lsb \onev_{N_{SQ}  \times  1 }, \zeros_{ N_{SQ} \times (\Nrx-1)} \rsb$ are identical.}
This receiver architecture encompasses those in Fig.~\ref{fig:channel model sign} and Fig.~\ref{fig:single antenna} as special
cases.
Figure \ref{fig:mulitiple antenna} again shows this model for $\Nrx=4$ and $N_{SQ}=3$.

\noindent
{\bf (d) Linear combining and multilevel quantization:} Corresponds to the set of all possible choices of $\Vv$ and $\tv$.

\section{Sign Quantization}
\label{sec:Known results}

The effect of quantization on the capacity of the MIMO channel  has been investigated thoroughly in the literature.
For conciseness, we review only the results on sign quantization of the channel outputs, corresponding to the architecture in Fig.~\ref{fig:channel model sign} for $N_{SQ}=\Nrx$,  which will be relevant in the remainder of the paper.
%

%Motivated by millimeter wave communication, various authors have studied the architecture i
%and a number of relevant results are available for this model.
%
The capacity of SISO channel with sign quantization of the outputs is attained by antipodal signaling.
\begin{lem}{\bf
%One-bit sign quantization SISO capacity
\cite[Th. 2]{singh2009limits}:}
\label{lem:one bit capacity}
The capacity of the SISO channel with sign quantization of the antenna output with $N_{SQ}=\Nrx$ is
\ea{
\Ccal_{\rm SISO}=1-H_2 \lb Q \lb \sqrt{P} \rb \rb.
\label{eq:one bit capacity}
}
\end{lem}
The capacity of the MISO channel with sign output quantization is obtained from the result in Lem. \ref{lem:one bit capacity}
by transforming this model into a SISO channel through transmitter beamforming, thus yielding
\ea{
\Ccal_{\rm MISO}=1-H_2 \lb Q \lb |\hv|\sqrt{P} \rb \rb.
\label{eq:only MISO}
}
For the SIMO and MIMO channel, capacity with sign quantization is known in the high-SNR regime.
\begin{lem}{\bf  \cite[Prop. 1]{mo2014high}.}
\label{lem:sigm  SIMO}
The capacity of the  SIMO  channel with sign quantization of the antenna output   with $N_{SQ}=\Nrx$ at high SNR satisfies
\ea{
 \log(\Nrx ) \leq \Ccal_{{\rm SIMO},a}^{\rm SNR \goes \infty} \leq  \log(\Nrx+1).
}
\end{lem}
\begin{lem}{\bf \cite[Prop. 3]{mo2015capacity}.}
\label{lem:finite capacity}
%{lem:high SNR MIMO}
%The high SNR capacity of the complex MIMO channel with one-bit sign quantization
The capacity of the  MIMO  channel with   sign quantization and $N_{SQ}=\Nrx$, and
for which $\Hv$ satisfies a \emph{general position} condition  (see \cite[Def. 1]{mo2015capacity}), is bounded at high SNR as
\ean{
\f 12 \log(K(N_{SQ},\Ntx) ) \leq \Ccal_{{\rm MIMO},a}^{\rm SNR \goes \infty} \leq  \f 12\log(K(N_{SQ},\Ntx)+1)
}
if $\Ntx<N_{SQ}$, where
\ea{
K(N_{SQ},\Ntx) = \sum_{k=0}^{2 \Ntx-1} {{2N_{SQ}-1}\choose{k}}.
}
If $\Ntx \geq N_{SQ}$, then $\Ccal_{{\rm MIMO},a}^{\rm SNR \goes \infty}   = N_{SQ}$.
%\ea{
%\Ccal_{MIMO,a}^{\rm SNR \goes \infty}   = N_{SQ}.
%}
\end{lem}
At finite SNR, upper and lower bounds on the capacity of the MIMO channel with sign quantization are known but are not tight in general~\cite[Sec. V.A]{mo2015capacity}.
%
%can be obtained by having the transmitter invert the channel matrix.
%%
%An upper bound for this regime can be obtained from the fact that $H_2(Q(x))$ is a decreasing convex function.
%%
%\begin{lem} {\bf  MIMO channel with  sign quantization \cite[Sec. V.A]{mo2015capacity}.}
%\label{lem:finite capacity}
%The capacity of a MIMO channel with output sign quantization with $\Ntx \geq \Nrx=N_{SQ}$ is be bounded  as
%\ea{
% & N_{R}\lb1-H_2\lb Q \lb \sqrt{ \f {P}{\trace \lb \Kv \rb} } \rb \rb \rb \leq \Ccal_{\text{MIMO}, a} \nonumber \\
% & \quad \quad \leq  N_{R}\lb1-H_2\lb Q \lb \sqrt{P \la_{\max}} \rb \rb \rb,
%%\f 12  \log \lb  \min\{\Nrx, \det \lb I+ H H^T\rb \}\rb
%}
%where $\Kv=(\Hv \Hv^T)^{-1}$ and  $\la_{\max}$ is the largest eigenvalue of $\Kv$.
%\end{lem}
%

\section{Main Results}
\label{sec:Main Results}
We begin by considering the capacity of the SISO channel for the receiver architectures in Sec. \ref{sec:Relevant Architectures}.
Capacity for the architecture~(a) is provided in Lem. \ref{lem:one bit capacity} (necessarily $N_{SQ}=1$) while the architectures  (b), (c) and (d) all correspond to the same model in which the channel output is quantized
through an $(N_{SQ}+1)$-level quantizer.
The capacity for this latter model can be bounded to within a small additive gap as shown in the next proposition.
%
%The capacity of the SISO channel with quantization constraint
%
\begin{prop}
\label{prop:multilevel SISO}
The capacity of the SISO channel with multi-level output quantization, $N_{SQ}>1$,
%, $P>1$
is upper-bounded as
\ea{
%
%\f 1 2 \log(\min\{P+1,M^2\})- H_2(P_e) - \f {P_e } 2 \log (\min\{P+1,M^2\})  \leq
\Ccal_{{\rm SISO}} \leq    \f 12 \log\lb \min \lcb P+1,(N_{SQ}+1)^2 \rcb \rb,
\label{eq:multilevel SISO}
}
and capacity is to within  $1$ bits-per-channel-use ($\bpcu$) from the upper bound in \eqref{eq:multilevel SISO}.
\end{prop}%
\begin{IEEEproof}
The upper bound \eqref{eq:multilevel SISO} is the  minimum between the capacity of the model without quantization constraints
and the capacity of the channel without additive noise.
%
%In the following we refer to this upper bound as the \emph{trivial upper bound}.
%
For the achievability proof, the input is chosen as an equiprobable $M$-PAM signal for
\ea{
M= \min \lcb   \lfloor  \sqrt{P} \rfloor , N_{SQ} +1 \rcb,
%>2,
}
in which the distance between the constellation points is
%0
such that the power constraint is met with equality.
At the receiver, the quantization thresholds are selected as the midpoints of the $M$-PAM constellation points.
The full proof is in  App. \ref{app:multilevel SISO}.
\end{IEEEproof}
For the SIMO and MIMO cases, given the generality of the formulation in \eqref{eq:max c}, rather than attempting to find the exact capacity $C(\cal{F})$ for each architecture in Sec. \ref{sec:Channel Model},
%
%The problem formulation in \eqref{eq:max c} is  very general so that an exact solution can be determined only for some specific choices of
%the analog processing constraints.
%
%For this reason,
we instead focus on approximate characterization  in the spirit of Prop.~\ref{prop:multilevel SISO},
%more specifically:
that is:
(i) the upper bound  is obtained  as the minimum among   two simple upper bounds %(in the following we refer to this upper bound as the \emph{trivial upper bound}),
and (ii) the achievability proof relies on a transmission scheme whose performance can be easily compared to the upper bound
to show a small gap  between the two bounds.
%expression,
%  so that the exact capacity  is bounded  to within the small gap  between these two bounds.
%
This approach provides an approximate characterization of capacity which is useful in comparing the performance of different architectures.
In the following, we extend the result in Prop.~\ref{prop:multilevel SISO} to the SIMO and MIMO cases.\footnote{Note that the MISO case follows from the SISO case as in \eqref{eq:only MISO}.}

\subsubsection{SIMO case}
The capacity for the architecture~(a) is obtained by selecting the antenna with the largest gain; for the architecture~(b) the capacity
%in Fig.~\ref{fig:single antenna}
is a rather straight-forward extension of the result in
Prop. \ref{prop:multilevel SISO}.
\begin{prop}
\label{prop:single antenna selection}
The capacity of the SIMO channel with single antenna selection and multilevel quantization is upper-bounded as
\ea{
\Ccal_{{\rm SIMO}, b} \leq \f 12 \log \lb \min \lcb 1+h_{\max}^2 P, (N_{SQ}+1)^2 \rcb \rb,
\label{eq:single antenna selection}
}
where $h_{\max}=\max_i h_i$ and the upper bound in \eqref{eq:single antenna selection} can be attained to within $1/2 \ \bpcu$.
\end{prop}
\begin{IEEEproof}
%The proof follows simply from the fact that only one antenna output can be selected for quantization.
The proof is provided in App. \ref{app:single antenna selection}
\end{IEEEproof}
For the architecture (c), sampling  more antennas allows the
receiver to collect more information on the input but  reduces
the  number of samples that can be acquired from each antenna.

\begin{prop}
\label{prop:simo multipe antenna}
The capacity of the SIMO channel with multiple antenna selection and multilevel quantization
for
%which $h_i>1 \ \forall i\in[1...\Nrx]$ and
%$|\hv|^2 P > 4$,
$P>\log(N_{SQ})>2$ and $h_i^2 >1$  is bounded as
{
%\small
\eas{
%& \Ccal_{{\rm SIMO}, c} \geq  \label{eq:achievable simo multiple} \\
%&  \quad \max_K \f 12 \log  \lb \min \lcb 1+ |\hv^{(K)}|_2^2  P,  \lb \f {N_{SQ}} K -1 \rb^2 \rcb \rb-2, \nonumber
%
& \max_K \f 12 \log  \lb \min \lcb 1+ |\hv^{(K)}|_2^2  P,  \lb \f {N_{SQ}} K +1 \rb^2 \rcb \rb-2
\label{eq:achievable simo multiple 1} \\
&  \quad \quad \leq  \Ccal_{{\rm SIMO}, c}  \leq \f 12 \log  \lb 1+|\hv|_2^2P, (N_{SQ}+1)^2 \rb,
\label{eq:achievable simo multiple 2}
}{\label{eq:achievable simo multiple}}}
where $\hv^{(K)}$  is the vector of the $K$ largest channel gains.
%the $K$ largest channel gains and $K\in [1 \ldots \Nrx]$.
%where the maximum is over $K$ such that
\end{prop}
\begin{IEEEproof}
The upper bound is derived similarly to Prop. \ref{prop:multilevel SISO}.
The achievable rate with finite uniform output quantization is  related to the achievable rate with infinite uniform output
quantization by bounding the largest difference between these two quantities under the conditions $P>\log(N_{SQ})$ and   \mbox{$h_i^2 P>1$}.
%the two values are to within a constant distance for all channel parameters.
%
In the model with infinite output quantization, a dither can be used to make the quantization noise
independent of the channel input and of the additive noise,
%
%in this model the dither renders the quantization noise uniform over the quantization interval and uncorrelated with the channel input and the additive noise.
%
%Successively we evaluate the attainable rate for the infinite uniform quantization channel:
%
so that the worst additive noise lemma
%from \cite{diggavi2001worst}
may then be used to  lower bound
the attainable rate as in \eqref{eq:achievable simo multiple}.
The full proof is provided in App. \ref{app:simo multipe antenna}.
\end{IEEEproof}

\begin{prop}
\label{prop:simo linear}
The capacity of the SIMO channel with linear combining and multilevel quantization is upper-bounded as
\ea{
\Ccal_{{\rm SIMO}, d} \leq \f 12 \log  \lb \min \lcb 1+|\hv|_2^2P, (N_{SQ}+1)^2 \rnone\rb,
\label{eq:simo linear}
}
and the upper bound in \eqref{eq:simo linear} can be attained to within $1/2 \ \bpcu$.
\end{prop}
\begin{IEEEproof}
With this architecture, the maximal ratio combining at the receiver results in the equivalent SISO channel with channel gain $|\hv|_2$.
The result in Prop. \ref{prop:multilevel SISO} can then be used to obtain the approximate capacity.
%
%The proof follows from the trivial upper bound
%and the achivability proof in Prop. \ref{prop:simo multipe antenna} for $K=\Nrx$.
%
%The full proof is provided in App. \ref{app:simo linear}.
\end{IEEEproof}
The results in Prop. \ref{prop:single antenna selection}, Prop. \ref{prop:simo multipe antenna} and Prop. \ref{prop:simo linear} are related as follows.
The results for the architecture (a) in Lem. \ref{lem:sigm  SIMO} and the architecture (b) in Prop. \ref{prop:single antenna selection}
show that the two architectures yield the same high-SNR behaviour when $\Nrx \geq N_{SQ}$.
When $\Nrx < N_{SQ}$, though, the architecture in (b) can attain higher performance at high SNR.
The architectures (c) and (d) differ as follows: in the former, the estimate of the transmitted message is implicitly obtained by combining the quantized information while, in the latter, combining occurs before quantization.
From Prop.~\ref{prop:simo multipe antenna} we gather the conditions under which  combining after quantization roughly attains the same performance
as combining before quantization: this occurs when the number of quantizers is sufficiently large so that the first term in the minimum in \eqref{eq:achievable simo multiple 1} dominates the channel performance.
\begin{prop}
\label{prop:simo multi}
The capacity of the SIMO channel with multiple antenna selection and multilevel quantization
is upper-bounded as
\ea{
\Ccal_{{\rm SIMO}, c} \leq \f 12 \log \lb 1+|\hv|_2^2P \rb,
\label{eq:upper trivial}
}
and the upper bound in \eqref{eq:upper trivial} can be attained to within $1 \ \bpcu$
when $N_{SQ}> \Nrx \sqrt{|\hv|_2^2 P+1}$ and  $h_i^2 >1$.
\end{prop}
\begin{IEEEproof}
%The proof is provided in App. \ref{app:simo multi}.
Under these assumption, the minimum in \eqref{eq:achievable simo multiple 1}  is attained by setting $K=N_r$, in which case the trivial outer bound of \eqref{eq:achievable simo multiple 2} can be attained to within $2 \bpcu$.
\end{IEEEproof}
%\markL{LUCA: Are you sure the formula in (17) is correct?}
%Note that the condition $h_i^2 P>1$ in Prop. \ref{prop:simo multi}
%

\subsubsection{MIMO case}
%
%The results in Prep. \ref{prop:single antenna selection} naturally extend to the MIMO case.
%
For the architecture (a),  inner and outer bounds are derived in \cite[Sec. V.A]{mo2015capacity}; for the architecture (b), an upper bound is derived in the next proposition.
\begin{prop}
\label{prop:mimo single}
The capacity of the MIMO channel with single antenna selection and multilevel quantization is upper-bounded as
\ea{ \label{eq:mimo_b}
\Ccal_{{\rm MIMO},b}\leq \f 12 \log \lb \min \lcb 1+|\hv_{\max}^T|_2^2P,(N_{SQ}+1)^2 \rcb \rb,
}
where $\hv_{\max}^T$ is the row of $\Hv$ with the largest norm and the upper bound in~\eqref{eq:mimo_b}
can be attained to within $2 \ \bpcu$.
\end{prop}
\begin{IEEEproof}
The proof is provided in App. \ref{app:mimo single}.
\end{IEEEproof}
%When multiple antennas can be selected, an inner bound can be derived as a variation of the water-filling
%For the architecture most general receiver architecture where any linear combination attains a maximal rate which is an interesting variation of the waterfilling solution.
%
%
%\newpage
%\onecolumn
%For the MIMO channel with multiple antenna selection, as in the architectures (c) and (d), an interesting achievable
% rate can be derived as  a variation of the
%classic water-filling solution.
%
For the architecture (d), the approximate capacity can be obtained as a variation of the classic water-filling solution.
%classic water-filling solution.
By decomposing the channel matrix through singular value decomposition, the channel can be transformed in  $K=\min \{N_t,N_r\}$ parallel channel with gains
 $\{\la_i\}$.
%
%
%Consider the case in which we wish to maximize the capacity of $K$ parallel SISO channels as in Prop. \ref{prop:multilevel SISO},
%which corresponds to the MIMO model in which $\Hv$ is a square matrix with elements $\lav$ on its diagonal (later, $\la$ are chosen to be the singular values of the channel matrix $\Hv$ so that $K=\min \{N_t,N_r\}$).
%under a power and quantization constraints:
Capacity is then obtained as
\ea{
\max \ \sum_{i=1}^K  \f 12 \log \lb \min \lcb 1+\la_i^2 P_i, ( N_{SQ,i}+1)^2\rcb \rb,
\label{eq:maximization}
}
where the maximization is over $P_i \in \Rbb^+, \ \sum_i P_i = P$, $N_{SQ,i} \in \Nbb, \ \sum_i N_{SQ,i} =N_{SQ}$ and $K \in [0,\min \{N_t,N_r\}]$.
By relaxing the integer constraint on the parameters $N_{SQ,i}$, we obtain to the outer bound
\small \ea{
& \Ccal \leq R^\star(\lav,P,N_{SQ})= \nonumber \\
& \lcb \p{
\sum_{i=1}^{\min\{\Nrx,\Ntx\}}\f 12 \log (1+\la_i P_i)  \\
\quad \quad \quad  \rm{if} \ \  \sum_{i=1}^{\min\{\Nrx,\Ntx\}} \lb \sqrt{1+\la_i P_i}-1 \rb \leq N_{SQ} \\
 K  \log \lb \f {N_{SQ}} K  +1 \rb\\
  \quad \quad \quad   {\rm \small otherwise},
}\rnone
\label{eq:general waterfilling}
}\normalsize
%\markL{LUCA: the first parameter of $R^\star$ is $\hv$ or $\lav$?}
%FIXED!
where $P_i$ are chosen as $P_i=(\mu-\la_i^{-2})^+$  and $\mu$ is the smallest value for which  $\sum_i P_i=P$  and
$K=\sum_i 1_{\{P_i>0\}}$.
The approximate capacity for the architecture (d) is obtained by showing that a rate sufficiently close to \eqref{eq:general waterfilling} is achievable.
%
%\markL{LUCA: Why does the indicator function depend only on $P_i$?}
% FIXED!
%
The capacity approaching transmission strategy is interpreted as follows: the classic water-filling
solution is approximatively optimal as long as each channel output can be quantized using  $N_{SQ,i} \approx \sqrt{1+\la_i P_i}-1$ quantizers. %\markL{LUCA: Maybe here we need an $\approx$ symbol, or a floor function to cast the integer number.}
% DONE BOSS!
If this condition is not satisfied, then the optimal solution is to uniformly assign the quantizers to all the active antennas. This leads to the next proposition.
\begin{prop}
%{\bf MIMO channel with linear combining and multilevel quantization.}
\label{prop:mimo linear}
The capacity of a MIMO channel with linear combining and multilevel quantization is upper-bounded as
\ea{
& \Ccal_{{\rm MIMO},d} \leq  R^\star(\lav,P,N_{SQ}),
% \min \lcb \Ccal_{\rm MIMO}(\min\{\Ntx,\Nrx\}),    K \log \lb \f {N_{SQ}} K +1 \rb  \rcb,
\label{eq:mimo linear}
}
and capacity  is to within a gap of  $3/2 K \ \bpcu$ from the upper bound in \eqref{eq:mimo linear} for $R^\star(\lav,P,N_{SQ})$ and $K$ in \eqref{eq:general waterfilling}.
 \end{prop}
 \begin{IEEEproof}
The proof is provided in App. \ref{app:mimo linear}.
\end{IEEEproof}
%Note that the gap from capacity in Prop. \ref{prop:mimo linear} grows as the number of active channels in the water-filling solution.
%
The result in Prop. \ref{prop:mimo linear} shows that sign quantization is sufficient to attain the optimal performance in the high SNR regime since $K=N_{SQ}$ yields the largest rate in \eqref{eq:general waterfilling} when $P \goes \infty$.
This follows from the fact that sign quantization, among all possible architectures, yields the largest number of information bits.
The optimality of this solution arises from the fact that the number of sign quantizer is  a fixed resource that limits, at the receiver side, the largest attainable rate.
%%%

%\lipsum[1-5]
\begin{figure*}%[b]
\vspace{-1.5 cm}
\hspace*{-1 cm}
\captionsetup[subfigure]{justification=centering}
 \begin{subfigure}[c]{0.35 \textwidth}
 \vspace{+.31 cm}
\resizebox{10.8 cm }{!}{
\begin{tikzpicture}
\node at (2.8,0)
{\includegraphics[trim={12.3cm 0 0 0},,clip=true,scale=.48 ]{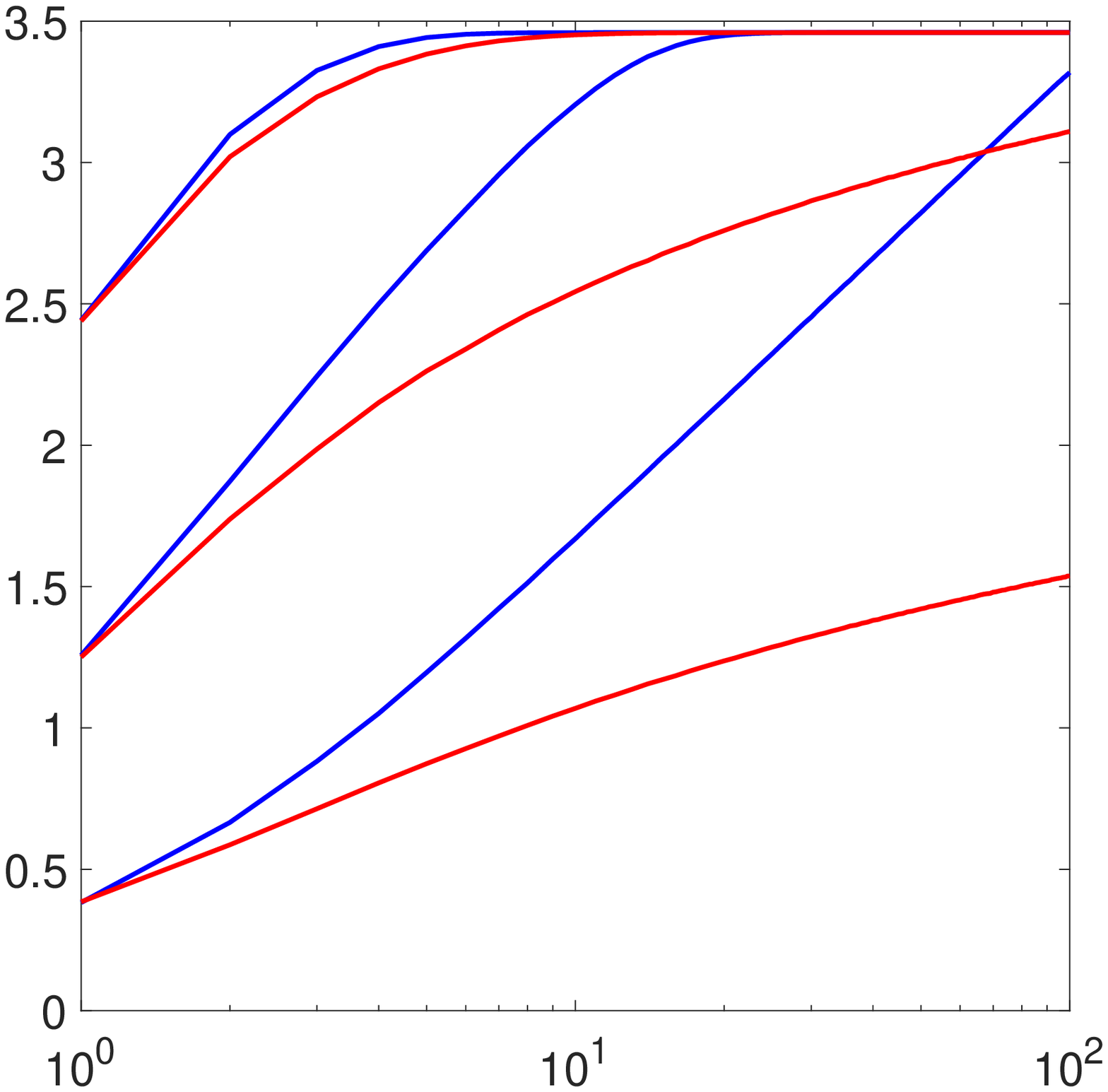}};
%{\includegraphics[trim=0cm 0cm 0cm 0cm,  ,clip=true,scale=.5 ]{FIGS/SIMOplot1Fig1}};
\node[rotate=90] at (-4.5,0.3) { {$R~[\bpcu]$}} ;
\node at (0,-4.5) {$\Nrx$};
\node at (-2,-3) {  \color{blue} {$\Ccal_{\rm SIMO,d}$}};
\node at (+3,-3) {  \color{red} {$\Ccal_{\rm SIMO,b}$}};
\draw[line width=.5 pt] (-1.8-.8,-3.1) -- (-1.8-.8,2.8);
\node at (-1.8-.8,-2.35) {$\bullet$};
\node at (-1.8-.8,+0.1) {$\bullet$};
\node at (-1.8-.8,2.7) {$\bullet$};
\draw[line width=.5 pt] (+2.8+0.5,-2.8) -- (-2.8+.5,2.8);
\node at (+0.95+0.5,-0.95) {$\bullet$};
\node at (-1.25+.5,+1.25) {$\bullet$};
\node at (-2.8+.5,2.8) {$\bullet$};
\node at (+1.2+0.5,1-0.95) {\small $P=1$};
\node at (+0.5,2.5) {\small $P=10$};
\node at (-2.7,3.5) {\small $P=100$};
\end{tikzpicture}
}
%\vspace{-1.5cm}
\caption{Prop.~\ref{prop:single antenna selection} vs. Prop.~\ref{prop:simo linear} for the SIMO channel with architectures (b) and (d) with $N_{SQ}=10$, $\Nrx\in [1 \ldots 10^2]$ and $P\in [1,10,100]$}
\label{fig:SIMO 1}
 \end{subfigure}
\begin{subfigure}[c]{0.35 \textwidth}
 \resizebox{10.3 cm }{!}{
\begin{tikzpicture}
%\node at (0,0) {\includegraphics[trim=0cm 0cm 0cm 0cm,  ,clip=true,scale=.5 ]{FIGS/SIMOplot1Fig2}};
\node at (2.8,0)
{\includegraphics[trim={12.3cm 0 0 0},,clip=true,scale=.48 ]{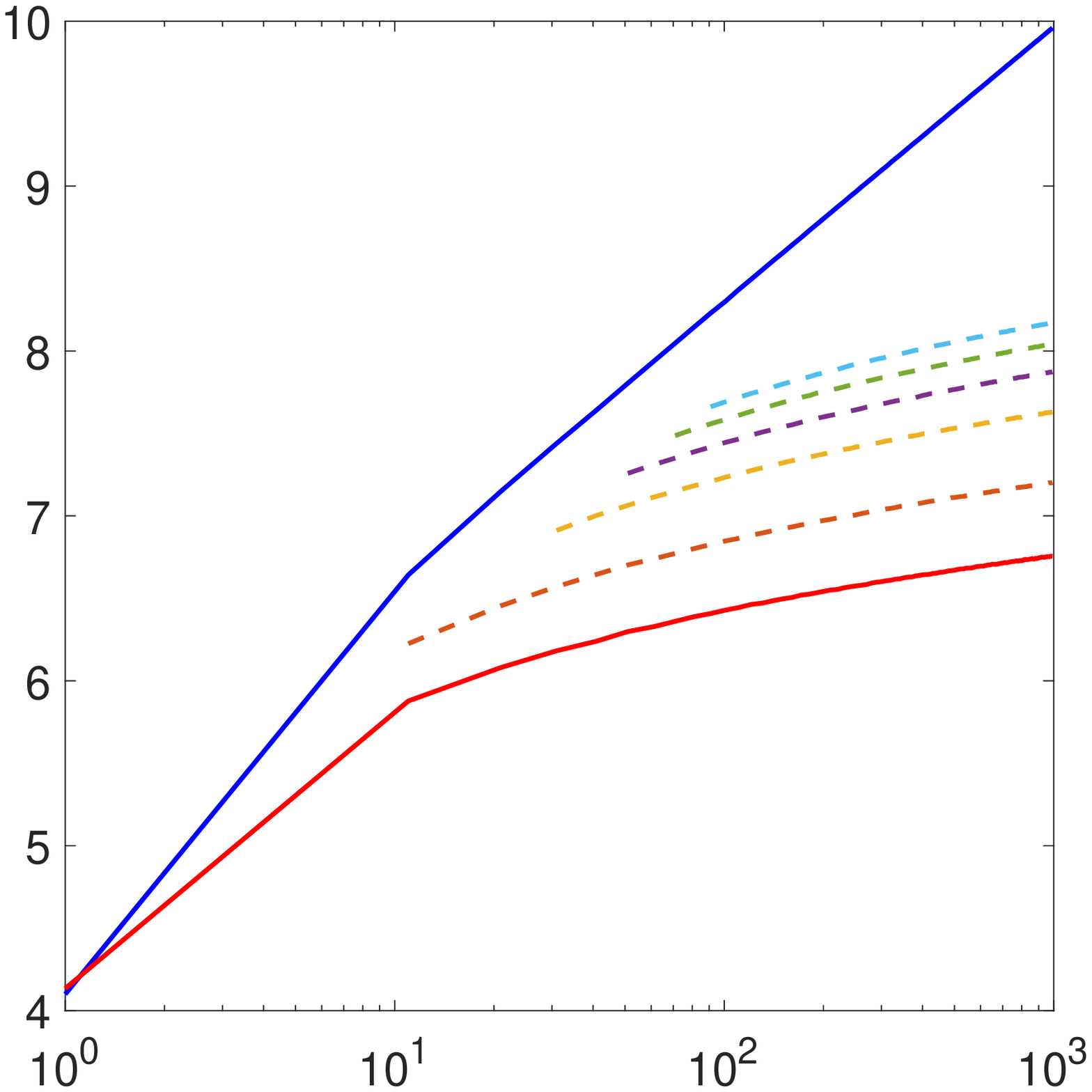}};
\node[rotate=90] at (-4.2,0.3) { {$R~[\bpcu]$}} ;
\node at (0,-4.5) {$\Nrx$};
\node at (-2.5,-1.25) {  \color{blue} {$\Ccal_{\rm SIMO,d}$}};
\node at (-1.5,-3) {  \color{red} {$\Ccal_{\rm SIMO,b}$}};
\node at (1,-3) { \textcolor[rgb]{0.00,0.59,0.00}{$\Ccal_{\rm SIMO,c}$}};
\node at (1.5,-2.5) { \textcolor[rgb]{0.00,0.59,0.00}{\small $K=2$}};
\draw[line width=.5 pt] (1,-2.5)  -- (1,-0.2);
\node at (1,-0.2) {$\bullet$};
\node at (2,-2) { \textcolor[rgb]{0.00,0.59,0.00}{\small $4$}};
\draw[line width=.5 pt] (1.5,-2)  -- (1.5,0.4);
\node at (1.5,0.4) {$\bullet$};
\node at (2.5,-1.5) { \textcolor[rgb]{0.00,0.59,0.00}{\small $6$}};
\draw[line width=.5 pt] (2,-1.5)  -- (2,0.8);
\node at (2,0.8) {$\bullet$};
\node at (3,-1) { \textcolor[rgb]{0.00,0.59,0.00}{\small $8$} };
\draw[line width=.5 pt] (2.5,-1)  -- (2.5,1.1);
\node at (2.5,1.1) {$\bullet$};
\node at (3.5,-.5) [fill=white]{ \textcolor[rgb]{0.00,0.59,0.00}{\small $10$}};
\draw[line width=.5 pt] (3,-.5)  -- (3,1.4);
\node at (3,1.4) {$\bullet$};
\end{tikzpicture}
}
% \vspace{-1cm}
 \caption{ Prop.~\ref{prop:single antenna selection} vs. Prop.~\ref{prop:simo linear} vs. Prop.~\ref{prop:simo multipe antenna} for the SIMO channel with architectures (b), (c) and (d) with $N_{SQ}=100$, $P=10^3$ and  $\Nrx\in [1 \ldots 10^3]$.}
 \label{fig:SIMO 2}
  \end{subfigure}
  \begin{subfigure}[c]{0.35 \textwidth}
   \vspace{+.3 cm}
\resizebox{10.3 cm }{!}{
\begin{tikzpicture}

%\node at (0,0) {\includegraphics[trim=0cm 0cm 0cm 0cm,  ,clip=true,scale=.5 ]{FIGS/MIMOplot2FigV3}};
\node at (2.8,0)
{\includegraphics[trim={12.3cm 0 0 0},,clip=true,scale=.48 ]{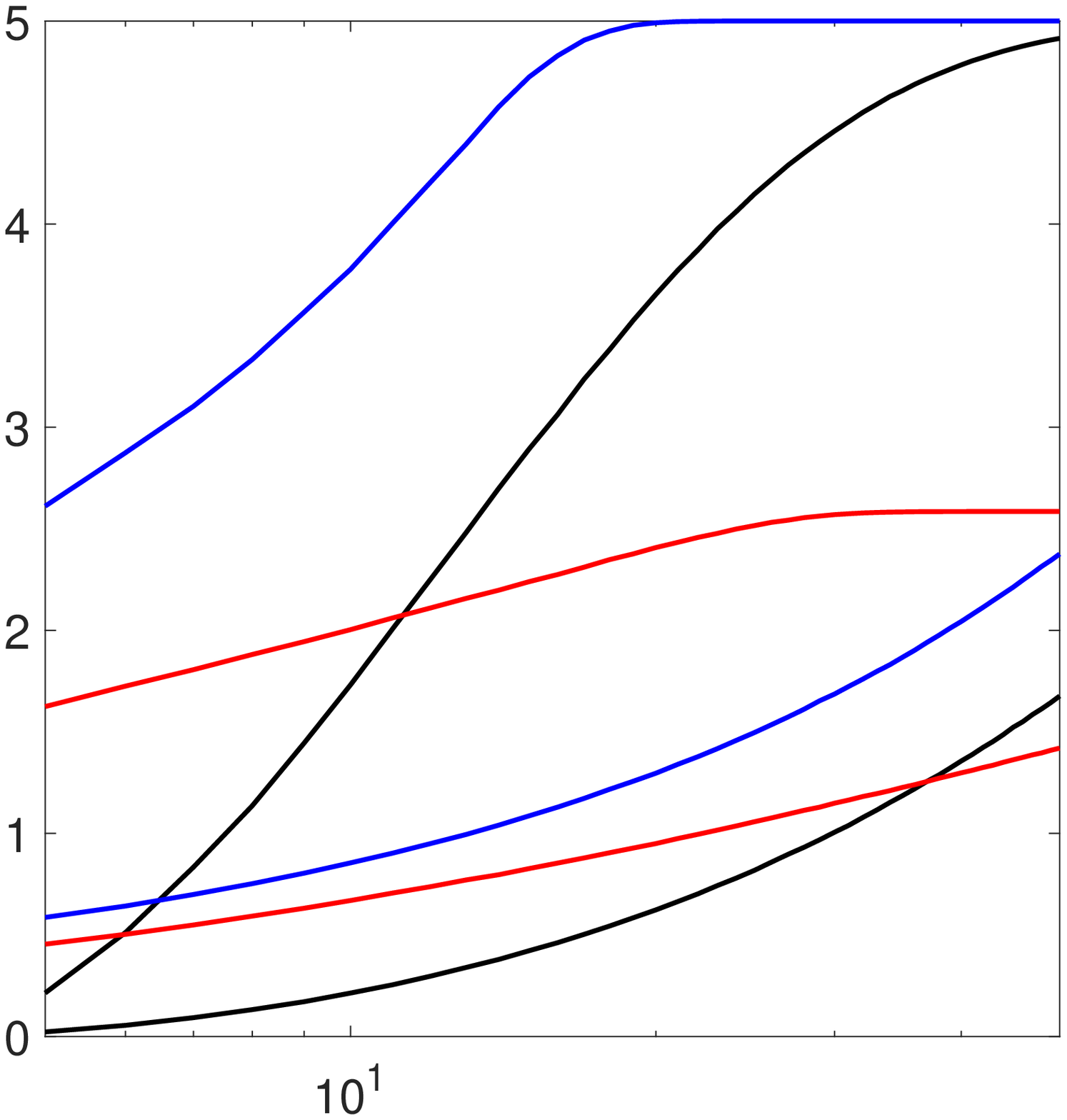}};

\node[rotate=90] at (-4.25,0.3) { {$R~[\bpcu]$}} ;
\node at (0,-4.5) {$\Nrx$};
\node at (-2.5,+3) {  \color{blue} {$\Ccal_{\rm MIMO,d}$}};
\node at (+3,+2) {  \color{red} {$\Ccal_{\rm MIMO,b}$}};
\node at (-0,+2) {   {$\Ccal_{\rm MIMO,a}$}};
\draw[line width=.5 pt] (-1.8-1.3,-2.7) -- (-1.8-1.3,2.7);
\node at (-1.8-1.3,-2.6) {$\bullet$};
\node at (-1.8-1.3,+0.5) {$\bullet$};
\draw[line width=.5 pt] (+2.5,+1.7) -- (+2.5,-1.9);
\node at (+2.5,-1.75){$\bullet$};
\node at (+2.5,+0.25) {$\bullet$};
\draw[line width=.5 pt] (-0.5,+1.75) -- (-0.5,-3);
\node at (-0.5,-0) {$\bullet$};
\node at (-0.5,-3.05) {$\bullet$};
\node at (-2,+.5) {\small $P=1$};
\node at (+2.5,-3) {\small $P=0.1$};
\end{tikzpicture}
}
% \vspace{-1cm}
 \caption{ Lem.~\ref{lem:finite capacity} vs. Prop.~\ref{prop:mimo single} for the MIMO channel with architectures (a), (b) and (d) with $N_{SQ}=\Ntx=5$, $\Nrx\in [5 \ldots 50]$ and $P \in \{0.1,1\}$.}
 \label{fig:MIMO}
  \end{subfigure}
\caption{
Average performance comparison.
}
\label{fig:Average performance comparison}
\vspace{-.5 cm }
\end{figure*}
\section{Numerical Evaluations}
\label{sec:Numerical Evaluations}

In the following, we evaluate the results in Sec. \ref{sec:Main Results} by considering the expected value of capacity $\Ccal(\Fcal)$ in
\eqref{eq:max c} when the channel gains $H_{ij}$ are drawn from a Gaussian distribution with mean zero and variance one.
%\footnote{This expected value corresponds to
%the ergodic capacity  with
%channel state information at the transmitter and the receiver.
%}
%
We begin by numerically evaluating the performance for the SIMO channel with single antenna and multilevel quantization
selection in Prop.~\ref{prop:single antenna selection} and with linear combining in Prop.~\ref{prop:simo linear}.
Figure \ref{fig:SIMO 1}  shows the upper bound expressions in \eqref{eq:single antenna selection} and \eqref{eq:simo linear} as a function of the
number of receiver antennas $\Nrx$ and for a fixed transmit
power $P$ and number of sign quantizers $N_{SQ}$.
For $\Nrx=1$, the performance of the two architectures is the same as the SISO channel in Prop. \ref{prop:multilevel SISO}, while, when
 $\Nrx$ increases, the performance approaches $\log(N_{SQ}+1)$, albeit at a slower rate for the single antenna selection case.
As the power increases, the transition between these two regimes requires fewer  antennas.
Consequently, the performance loss of the receiver
architecture in Figure \ref{fig:single antenna}, in comparison with linear combining receiver, decreases as the transmit power grows large.

The performance of multiple antenna selection for the SIMO case is shown in Figure \ref{fig:SIMO 2}: in this figure, we plot
the upper bound in Prop.~\ref{prop:single antenna selection} and Prop.~\ref{prop:simo linear} together with those in Prop.~\ref{prop:simo multi}.
From Figure \ref{fig:SIMO 2} we observe how increasing the number of antennas that are selected impacts the achievable rate, reducing the gap from the performance of the architecture with linear combining and multilevel quantization.

The performance for the MIMO case is presented in Fig.~\ref{fig:MIMO}: in this figure, we show the performance difference
between the architectures (a) from \cite[Sec. V.A]{mo2015capacity}.
%\markL{LUCA: Which Lemma? To fix also in the caption of Figure 2c.} Prop. \ref{prop:mimo single}.
% as compared to Prop. \ref{prop:mimo linear}.
%
Single antenna selection with multilevel quantization performs well when the number of receive antennas is small but its performance is surpassed by multi-antenna selection and sign quantization
as the number of receiver antennas grows.
This follows from the fact that the attainable rate with single antenna selection converges to $\log(N_{SQ}+1)$ as $\Nrx$ grows while
sign quantization converges to $N_{SQ}$.
% in the lemma N_r=N_SQ.
%I wanted to have  a better and more general proof, but i had to cut things, so i jsut change n_r=N_SQ here :(
%\markL{LUCA: I don't get this result $N_{SQ}$. Is this the result of Lemma III.3 for $\Ntx<\Nrx$?}
%
It is interesting to observe that these two simple receiver architectures, together, are able to closely approach the performance in Prop. \ref{prop:mimo linear}.
\section{Conclusion}
\label{sec:Conclusion}

A general approach to model receiver architectures for MIMO channels with low-resolution output quantization has been proposed.
In our formulation, the antenna outputs undergo analog processing before being quantized using $N_{SQ}$ sign quantizers.
Analog  processing is embedded in the channel model description while the channel output corresponds to the output of the sign quantizers.
Through this formulation, it is then possible to optimize the capacity expression over the set of feasible analog processing operations while keeping the number of sign quantizers fixed.
%
%Consequently, the result of this optimization is the receiver architecture yielding the best description of the antenna outputs in $N_{SQ}$ bits under
%the given restrictions on the analog processing capabilities.
%
%We apply this approach to the study of MIMO channels under four analog receiver architectures: antenna selection and  output sign quantization, single antenna selection and multilevel quantization, multiple antenna selection and multilevel quantization, and linear combining and multilevel quantization.
%

\bibliographystyle{IEEEtran}
\bibliography{steBib1}

\newpage
\onecolumn
\appendices

\section{Proof of Prop. \ref{prop:multilevel SISO}}
\label{app:multilevel SISO}

\noindent
$\bullet$ \textbf{Converse:}
The capacity of the SISO channel with multilevel quantization is necessarily dominated by the capacity of the AWGN channel without quantization constraints and by the capacity of the channel with channel with output quantization but no additive noise.

The upper bound
\ea{
\Ccal_{{\rm SISO}} \leq    \f 12 \log\lb P+1\rb,
\label{eq:AWGN out}
}
is obtained as the capacity of the channel without quantization constraints.
The upper bound
\ea{
\Ccal_{{\rm SISO}} \leq     \log\lb N_{SQ}+1 \rb,
\label{eq:quantizer out}
}
is obtained as the capacity of the channel without additive noise.
The intersection of the outer bounds in \eqref{eq:AWGN out} and \eqref{eq:quantizer out} yields the outer bound in \eqref{eq:multilevel SISO}.
In the following we refer to this upper bound as the \emph{trivial upper bound} for brevity.

\smallskip
\noindent
$\bullet$ \textbf{Achievability:}
If $N_{SQ}=1$, then capacity is provided by Lem.~\ref{lem:finite capacity} for any  $P>0$.

Let us first consider the case is which  $P\leq 6$ and $N_{SQ}>1$: in this parameter regime
it can be verified through numerical evaluations that the capacity expression in \eqref{eq:one bit capacity} is to within $1/2 \ \bpcu$ from the infinite quantization capacity in \eqref{eq:AWGN out}.
This implies that the achievability proof in Lem. \ref{lem:finite capacity} is sufficient to show the approximate capacity in this parameter regime.

\medskip

For $P>6$ and $N_{SQ}\geq  2$, consider the achievable scheme in which the channel input is an equiprobable $M$-PAM constellation while, at the receiver,
the $M-1$ sign quantizers thresholds are chosen as the midpoints of the transmitted constellation points.
The parameter $M$ is chosen according to whether performance is limited by the transmit power or by the number of available quantizers.
When $1/2 \log(P+1) \geq \log(N_{SQ}+1)$, the number of available sign quantizers dominates the performance and $M$ is chosen as $N_{SQ}+1$, which is the largest number
of channel inputs that can be distinguished at the receiver.
When $\log(N_{SQ}+1) > 1/2 \log(P+1)$, then the available transmit power dominates the performance and $M$ is chosen as $\lfloor \sqrt{P} \rfloor$.
%
%
%For these reasons, define

Following these reasoning, we define
\ea{
M=\min \{N_{SQ}+1,\lfloor \sqrt{P} \rfloor \} \geq 3,
\label{eq:choose M}
}
and denote support of the input $X$ as
\ea{
\Xcal=\{x_1,\ldots ,x_M\},
\label{eq:M pam support}
}
for $\{x_m\}_1^M$ are in increasing order.
%
%
%\newpage
%
For $M$ is even, we choose $\Xcal$ as
\ea{
\Xcal
& =\De \cdot \lb  \lsb -M/2+1, \ldots ,+M/2  \rsb-1/2\rb,
\label{eq:M pam support even}
}
while, for $M$ odd, we let $\Xcal$ be equal to
\ea{
\Xcal= \De \lsb -\f {M-1} 2, \ldots, \f {M-1} 2 \rsb,
\label{eq:M pam support odd}
}
for
\ea{
\De & =\sqrt{\f{12 P}{M^2-1}}.
\label{eq:delta}
}
For $\Xcal$ in either \eqref{eq:M pam support even} or \eqref{eq:M pam support odd}, let the  channel input be uniformly distributed on the set $\Xcal$; note that, by construction, the power constraint is attained with equality, \emph{i.e.} $\Ebb [X^2]= P$.
%  $\Ebb[X]=0$ and $\Ebb [X^2]= P$.

At the receiver, the channel output is quantized using $M-1$ sign quantizers, each with threshold $t_k$ obtained as
\ea{
t_m= \f 12 \lb x_{m} + x_{m+1} \rb, \quad m \in [1,\ldots,M-1].
\label{eq:th siso}
}
Note that, by definition, $M-1\leq N_{SQ}$ so that the constraint on the number of available sing quantizers is respected.
In particular, for the case in which $N_{SQ}+1>\sqrt{P}$, we have that not all the sign quantizers are employed at the receiver.
In this scenario a better performance can be attained by employing all the available quantizer: for simplicity in the analysis, we only consider the sub-optimal strategy which employs $M-1$ of the  $N_{SQ}$ available quantizers.

\medskip

%Note that note all quantizers are necessary to approach the capacity to within a small additive gap.
%
For convenience of notation,
%
%and given the restriction of the feasible quantizer outputs in \eqref{eq:multilevel},
we express  $\Yv$ in \eqref{eq:quantized model} through the random variable $\Xh$ with support $\Xcal$ defined as
% \blue{LUCA: I don't understand this part: $\Yv$ is the sign quantizers output, while $\Xh$ is the symbol estimate. They represent different things. Why do you want to put them in relationship? I understand that in the quantization-limited regime they are closely related; but in the power-limited regime they are not. Maybe here you are thinking that $M-1 = N_{SQ}$, but this is not necessarily true: for example, when we are power limited and $N_{SQ}$ is very large.}
%
%STE: i added a comment to explain that we only use M-1 quantizer: this should make it clear that we have a one to one mapping!
%
\ea{
\Pr[\Xh=x_m]=\lcb
\p{
\Pr[W \leq t_1]   &  m=1  \\
\Pr[t_{m-1}<W \leq t_{m}] & m\in [2,\ldots, M-1] \\
\Pr[W > t_{M-1}]   &  m=M.
}
\rnone
\label{eq:xhat}
}
The mapping in \eqref{eq:xhat} is a one-to-one mapping since $\Yv_i$ is of the form
\ea{
\Yv_i=[\underbrace{-1 \ldots -1}_{M^{-}}, \underbrace{+1, \ldots +1}_{M^{+}}]^T,
\label{eq:multilevel}
}
with $M^{-},M^{+} \geq 0$ and  $M^{-}+M^{+}=M$,  so that the $M-1$ sign quantizer outputs have a one-to-one correspondence with $M$ possible values of $\Xh$.
%
%\blue{LUCA: Again, see previous comment.}
% I think we addressed this
%

With the definition in \eqref{eq:xhat} and for the channel input uniformly distributed over the support in \eqref{eq:M pam support even} and \eqref{eq:M pam support odd},
we obtain the inner bound
\ea{
R^{\rm IN}
& = H(\Xh)-H(\Xh|X),
\label{eq:inner bound xh}
}
where
\eas{
\Pr[\Xh=\xh|X=x] & = \Pr\lsb |Z-(\xh-x)|< \f {\De} 2\rsb \\
\Pr[\Xh=\xh]& =\f 1 M\sum_{m=1}^{M} \Pr\lsb \labs Z-(\xh-x)\rabs< \f \De 2\rsb,
}
where $Z \sim \Ncal(0,1)$ and $x,\xh \in \Xcal$.

The entropy term $H(\Xh)$ in \eqref{eq:inner bound xh} is lower-bounded as
\ea{
H(\Xh) \geq  M \min_{\xh \in \Xcal}   -P_{\Xh}(\xh)  \log P_{\Xh}(\xh),
\label{eq:H xh min}
}
and, given the symmetry in the input constellation,
we have that the minimum $P_{\Xh}(\xh)$ is obtained at $\xh=\pm \De/2$ for $M$ even,  and at $\xh=0$ for  $M$ odd.
Note moreover that, the minimum $P_{\Xh}(\xh)$ is at most $1/M\le 1/3 < e^{-1}$: for $x<e^{-1}$, the function $-x \log(x)$ is a positive increasing in $x$, so that a lower bound
on $P_{\Xh}(\xh)$ produces a lower bound to the RHS of \eqref{eq:H xh min}.
%
%
%\newpage
%
% Also, we note that for $M\ge 3$, the minimum $P_{\Xh}(\xh)$ is at most $1/M\le 1/3$. For $P_{\Xh}(\xh)\le 1/3$ the function $P_{\Xh}(\xh)  \log P_{\Xh}(\xh)$ is decreasing in $P_{\Xh}(\xh)$, so the function is maximized by the minimum $P_{\Xh}(\xh)$.
%
%For the case of $M$ even we write
%\ea{
%t_k = \f 12 \lb x_{k} + x_{k+1} \rb, \quad k \in [1,\ldots,M-1].
%}%\blue{LUCA: I modified (31a)-(31d) as follows. Now the problem is that the last inequality does not seem correct, due to Jensen's ineq and convexity of Q function.}
%
For this reason, when $M$ is even,  we lower bound $P_{\Xh}(+\De/2)=P_{\Xh}(-\De/2)$ as
\ea{
P_{\Xh}(+\De/2)
%& = \f 1 M \sum_{k=0}^M \Pr \lsb \Xh= \De/2 | X=x_k \rsb \\
&  = \f 1 {M} \lb \lb  1-2 Q(\De/2)\rb+ \sum_{k=2}^{+M/2} \lb Q((k-2)\De+\De/2)-Q((k-1)\De+\De/2) \rb \rnone  +  \nonumber\\
& \quad \quad \lnone  +\sum_{k=+1}^{+M/2} \lb Q((k-1) \De+\De/2)-Q( k \De+\De/2) \rb  \rb \nonumber \\
& = \f 1 {M} \lb 1-2 Q(\De/2)+ \lb Q(\De/2)-Q((M-1)\De/2)\rb  +\lb Q(\De/2)-Q((M+1) \De/2)\rb \rb \nonumber\\
& =\f 1 {M}  \lb 1-Q((M-1)\De/2)-Q((M+1) \De/2) \rb \nonumber\\
& \geq  \f 1 {M}  \lb 1-2Q((M-1) \De/2) \rb.
\label{eq:bound pxh M even}
}
%{\label{eq:bound HXh}}
%
Similarly, for the case of $M$ odd, we have
\ea{
P_{\Xh}(0)
%& = \f 1 M \sum_{k=0}^M \Pr \lsb \Xh= \De/2 | X=x_k \rsb \\
&  = \f 1 {M} \lb \lb  1-2 Q(\De/2)\rb+ 2 \sum_{k=1}^{+(M-1)/2} \lb Q((k-1)\De+\De/2)-Q(k\De+\De/2) \rb \rb \nonumber \\
& =  \f 1 {M}  \lb 1-2Q( M \De/2)  \rb.
\label{eq:bound pxh M odd}
}
By plugging \eqref{eq:bound pxh M even} and \eqref{eq:bound pxh M odd} in \eqref{eq:H xh min}, depending on the value of $M$, we obtain the bound
\ea{
\min_{\xh \in \Xcal} P_{\Xh}(\xh) \geq  \f 1 {M}  \lb 1-2Q((M-1) \De/2) \rb.
\label{eq:H xh min 2}
}
%\newpage
Let $\Qt=Q((M-1) \De/2)$ for convenience of notation and further bound \eqref{eq:H xh min 2} as
\eas{
H(\Xh) & \geq - M \f 1 M   \lb 1-2\Qt \rb  \log  \lb \f 1 {M}  \lb 1-2\Qt \rb  \rb \nonumber \\
%
%&  =  \log  \lb \f  {M}  {1-2Q((M-1) \De/2)} \rb +2Q((M-1) \De/2) \log  \lb \f 1 {M}  \lb 1-2Q((M-1) \De/2) \rb  \rb\\
& = \log M  - (1-2\Qt)\log(1-2 \Qt)-2 \Qt \log(M) \nonumber \\
& \geq  \log M  -2 \Qt \log(M) \label{eq:term 1} \\
& \geq  \log M -0.2 \label{eq:term 2},
}
where \eqref{eq:term 1} follows from the fact that the function $-\log(1-2 \Qt)-2 \Qt \log   (1-2 \Qt)$ is positive defined while
\eqref{eq:term 2} from the bound
\eas{
 \Qt& = Q \lb  \f 1 2 {(M-1) \sqrt{ \f {12 P}{ M^2-1}}} \rb \\
 & = Q  \lb \sqrt{\f {(M-1)^2}{M^2-1}} \sqrt{3 P} \rb  \\
 \label{eq:cond 1}
 & \leq Q(\sqrt{3 P}),
 %& \geq Q(\sqrt{6})
 %P-\sqrt{P} \rb
}
so that
\ea{
2\Qt\log(M)  \leq 2 Q(\sqrt{3 P}) \log(\sqrt{P}) & \leq  0.02,
\label{eq:cond 2}
}
where \eqref{eq:cond 2} follows from the fact that $Q(\sqrt{3 P}) \log(\sqrt{P}) $ is a decreasing function for $P>6$.

Accordingly, we conclude that
\ea{
H(\Xh) \geq \log M -0.02.
\label{eq:eq:bound 1X}
}

%
%\newpage
%
%a decreasing negative function of both $\Qt$ and $M$:
%\ea{
%\Qt \geq  \f 12 (P-\sqrt{P}).
%\label{eq:low bound}
%}
%Together with $M\geq 3$, the bound in \eqref{eq:low bound} can be used to tighten \eqref{eq:term 2} as
%\ea{
%H(\Xh) \geq \log M -0.2
%\label{eq:{eq:bound 1}}
%}
%
%
%
%\newpage
%
%where \eqref{eq:term 0}  is obtained as follows: $P<M^2 \De^2<4 P$ for $M\geq 2$ so that
%\ea{
%\f 1 2 (M-1)\De \geq  \f 12 (P-\sqrt{P}).
%\label{eq:low bound}
%}
%The function $2Q(x) \log(1-2Q(x))$ is a increasing, negative defined function so that
%\ea{
%2Q((M-1) \De/2) \log   \lb 1-2Q((M-1) \De/2) \rb \geq -0.1747,
%}
%when $P>4$ as by assumption.
%%
%Similarly, \eqref{eq:term 1} follows from the bound in \eqref{eq:low bound}  to yield
%\ea{
%2Q((M-1)\De/2) \log(M) \leq 2Q((P-\sqrt{P})/2) \log(\sqrt{P}).
%}
%The function $2Q((P-\sqrt{P})/2) \log(\sqrt{P})$ is decreasing for $P>4$ so that
%\ea{
%2Q((P-\sqrt{P})/2) \log(\sqrt{P}) \leq 0.3173.
%}
%
%and the maximum in this interval is $0.32$
%
%\blue{LUCA: I don't get the 3rd inequality. There should be a sign error in the last term}
% SR: should be fixed now!
%
%
%\newpage
%
%%
%Note now that $P<M^2 \De^2<4 P$ for $M\geq 2$ so that,
%\ea{
%2Q((M-1)\De/2) \log(M) \leq 2Q((P-\sqrt{P})/2) \log(\sqrt{P}) \leq  -0.18,
%}
%where we have used the simple bound $Q(x)\le e^{-x^2/2}$.
%Finally, we obtain
%\ea{
%H(\Xh) \geq \log(M)-0.18,
%\label{eq:bound 1}
%}
%when $P>4$.
%%
%\newpage
Next, we wish to upper bound the entropy term $H(\Xh|X)$ in \eqref{eq:inner bound xh}.
Note that, for each $X=x_m$,  $H(\Xh|X=x_m)$,  corresponds to  the entropy of a Gaussian random variable with mean $x_m$ and unitary variance which is quantized with $M$-level uniform quantization of step $\De$.
From the ``grouping rule for entropy'' \cite[Prob. 2.27]{cover2006elements}
%\blue{LUCA: Check reference}
we have that the value of this entropy  is smaller than the entropy of a Gaussian variable
 with infinite uniform quantization of step $\De$.

Let us denote as $N^{\De}$ the infinite quantization of a Gaussian variable with step $\De$; more specifically,
 $N^{\De}$ is defined as the random variable with support $\Zbb$ and for which $\Pr[N^{\De}=z], \ z \in \Zbb$ is obtained as
%
%To bound the term, we then define the random variable $N^{\De}$ a discrete random variable with support $\Xcal^{\De}$ defined as
%
\eas{
\Pr[N^{\De}=0]   &= \Pr \lsb  - \f \De 2 \leq X <  +\f \De 2  \rsb  \\
\Pr[N^{\De}=k]& =\Pr \lsb  (k-1) \De+\f \De 2 \leq X <  k\De+ \f \De 2  \rsb, \quad k \in \Zbb \setminus \{0\}.
}{\label{eq:quantized gaussian}}
The entropy $H(N^{\De})$ can be expressed as
\eas{
&H(N^{\De}) = -\lb 1-2 Q(\De/2) \rb \log \lb 1-2 Q(\De/2) \rb
\label{eq:quantized gaussian de 1}\\
& \leq 0.15 - 2 \sum_{k=0}^{\infty} \lb  Q(k \De+\De/2 )-Q((k+1)\De+ \De/2) \rb
\log \lb Q(k \De+\De/2 )-Q((k+1)\De+ \De/2) \rb.
\label{eq:quantized gaussian de 2}
}
For  $\Delta$ in \eqref{eq:delta}, we necessarily have $\De>2\sqrt{3}$, and thus
\eas{
& Q(k \De+\De/2 )-Q((k+1)\De+ \De/2)  \\
&<  Q(\De/2)- Q(3\De/2) \nonumber\\
& < Q(\De/2) < e^{-1}.
}{ \label{eq:stuff less then}}
Using the bound in \eqref{eq:stuff less then}, together with the fact that $-x\log (x)$ is an increasing function of $x$ for $x\le e^{-1}$,
we have that an upper bound on the term
$Q(k \De+\De/2 )-Q((k+1)\De+ \De/2)$ results in an upper bound on the quantity in \eqref{eq:quantized gaussian de 2}.

Next, note that for $k>1$, we have
\ea{
Q(k \De+\De/2 )-Q((k+1)\De+ \De/2)
\leq Q(k \De)-Q(2 k \De) \leq e^{-\f 1 2 -k^2 \De^2}-e^{ -2 k^2 \De^2},
}
so that, by numerical integration methods, we obtain the bound
\ea{
& - 2 \sum_{k=1}^{\infty} \lb  Q(k \De+\De/2 )-Q((k+1)\De+ \De/2) \rb \nonumber \\
& \leq 0.03+  \int_{x=1}^{\infty} \lb e^{-\f 1 2 -k^2 \De^2}-e^{ -2 k^2 \De^2} \rb \diff x \leq 0.25.
\label{eq:key}
}
Plugging the bound \eqref{eq:key} in \eqref{eq:quantized gaussian de 2}
we obtain
\ea{
H(N^{\De}) \leq 0.15+0.25=0.4
\label{eq:bound 2}
}
Finally, combining \eqref{eq:eq:bound 1X} and \eqref{eq:bound 2}
\ea{
I(X;\Xh)\geq \log(M)-\f 12,
}
which is the desired result.
%
%\blue{LUCA: I couldn't reduce the gap below 0.5 \bpcu.}
% SR: just re-evaluated everything in one go, so that the bound is smaller
\label{}
\section{Proof of Prop. \ref{prop:single antenna selection}}
\label{app:single antenna selection}
When only one antenna can be selected, the result in Prop. \ref{prop:multilevel SISO} can be used to bound the capacity maximization in \eqref{eq:max c} to within $1/2 \ \bpcu$ from the trivial outer bound
\ea{
\Ccal(\Fcal) \leq \max_k \f 12 \log \lb 1+h_k^2 P, (N_{SQ}+1)^2\rb.
\label{eq:max sas1}
}
%since the capacity of each antenna choice can be bounded using Prop. \ref{prop:multilevel SISO}.
%
The function on the RHS of \eqref{eq:max sas1} is increasing in $k$ when $h_k$ are ordered in increasing order, thus yielding the desired result.

%
%\newpage
\section{Proof of Prop. \ref{prop:simo multipe antenna}}
\label{app:simo multipe antenna}

The outer bound in \eqref{eq:achievable simo multiple 2} is the trivial outer bound as defined in App. \ref{app:multilevel SISO} while the inner bound in \eqref{eq:achievable simo multiple 1} is derived in the following.
In the remainder of this appendix, the channel coefficients $h_i$ are taken positive: this assumption is without loss of optimality as the noise distribution is symmetric.
Also, in the following, we assume without loss of generality that the terms $h_k$ are in descending order.

\medskip
\noindent
{\bf Achievability:}
If $|\hv|_2^2 P \leq 15$ or  $N_{SQ}\leq 3$, then
\ea{
	& \f 12 \log  \lb \min \lcb 1+ |\hv^{(K)}|_2^2  P, \lb  N_{SQ}  +1 \rb^2 \rcb \rb  \nonumber \\
	& \leq  \quad \f 12 \log  \lb \min \lcb 1+ |\hv|_2^2  P, \lb  N_{SQ}  +1 \rb^2 \rcb \rb  \leq 2
	\label{eq:bound P small}
}
from which we conclude that \eqref{eq:achievable simo multiple 1} is less than zero in this parameter subset.
Since the rate zero is trivially achievable, the inequality in \eqref{eq:bound P small} proves that \eqref{eq:achievable simo multiple 1} is achievable.

\medskip

If $|\hv|_2^2 P > 15$ and  $N_{SQ}> 3$,  the achievability of the bound in \eqref{eq:achievable simo multiple 1} is shown by letting the channel input
 be the sum of an $M$-PAM signal plus a dither.
For this receiver architecture dithered quantization is necessary to evaluate the performance of the combining of the sampled channel outputs.

Similarly to \eqref{eq:choose M}, let us we define $M$ as
\ea{
M= \Bigl \lfloor \min \lcb \f {N_{SQ}} K,  |\hv^{(K)}|_2 \sqrt{P}  \rcb-1 \Bigr \rfloor.
\label{eq:choose M 2}
}

For $M$ in \eqref{eq:choose M 2}, note that
\ea{
\eqref{eq:achievable simo multiple 1}  &= \f 12 \log  \lb \min \lcb 1+ |\hv^{(K)}|_2^2  P,  \lb \f {N_{SQ}} K +1 \rb^2 \rcb \rb   \\
& \leq \log(M+2), \nonumber
}
so that when $M \leq 2$, the expression in  \eqref{eq:achievable simo multiple 1} is less than zero which is trivially achievable.

\medskip
For $M \geq 3$,  let the channel input be obtained as
\ea{
X=S+U.
\label{eq:input dither}
}
where $S$ is an $M$-PAM signal for $M$ in \eqref{eq:choose M 2},  with support as in \eqref{eq:M pam support even}  for $M$ even, or as \eqref{eq:M pam support odd}
for $M$ odd but where $\De$  is chosen as
\ea{
\De =\sqrt{ \f {12 \al P }{M^2-1} }.% \geq \f {2 \sqrt{3}}{|\hv^M|},
}
The variable $U$ in \eqref{eq:input dither} is quantization dither, that is $U \sim \Ucal([-\De/2,+\De/2])$ and $U \perp S$.
%
%\newpage
%
Since $\Ebb[U^2] = \De^2 /12$,  the power constraint is satisfied with equality by setting
\ea{
\al P = P-\f {\De^2} {12},
}
which yields
\ea{
\De^2=\f {12 P}{M^2}.
}
%Note also that
%\ea{
%\De^2 \geq \f { 12}{|\hv^{(K)}|_2^2}.
%}

At the receiver, the $K$ antennas with the best SNR are each quantized with an $(M+1)$-level quantizer.
%\newpage
More specifically, the $k^{\rm th}$ antenna output, $k \in [1 \ldots K]$, is quantized with thresholds  $t_{i}^{(k)}$ for $i\in [0 \ldots M]$  chosen as
\eas{
t_0^{(k)} & = h_i \lb  x_1-\f {\De} 2  \rb \\
t_{m}^{(k)}& =  \f {h_i} {2} \lb x_{m} + x_{m+1} \rb, \quad m \in [1,\ldots,M-1] \\
t_{M}^{(k)} & = h_i \lb  x_M+\f {\De} 2  \rb.
}{\label{eq:finite out quan}}
Note that, although channel input has $M$ possible values but the receiver uses an $(M+1)$-level quantizer to quantize each of the $K$ best antenna outputs:
 two additional quantization levels are used to detect  whether the channel output is below $h_i(x_1-\De/2)$  or  above $h_i(x_M+\De/2)$
  (as specified at the beginning of the appendix,  the channel coefficients are assumed to be positive and with decreasing magnitude without loss of generality).
%
%
%units larger/smaller than the largest/smallest input value (scaled by $h_i$).
%%

Note that  the total number of quantizers employed at the receiver is $K (M+1) \leq   N_{SQ}$, so that the constraint on the total number of available sign quantizers
is satisfied.
As for the proof in App. \ref{app:multilevel SISO}, it is possible that not all the sign quantizer are utilized in this achievable scheme.

Next, similarly to \eqref{eq:xhat}, we define $\Xh^{(k)}$ for $k\in[1,\ldots,K]$ as
\ea{
\Pr[\Xh^{(k)}=x_m]=\lcb
\p{
\Pr[W_k \leq t_0^{(k)}]   &  m=0 \\
\Pr[t_{m-1}^{(k)}<W_k \leq t_{k}^{(k)}] & m \in [1,\ldots, M] \\
\Pr[W_k > t_{M}^{(k)}]   &  m=x_{M+1},
}
\rnone
\label{eq:xhat 2}
}
where $x_m$ for $m\in [1,M]$ is as in \eqref{eq:M pam support} while we additionally let $x_0=x_1- \De$ and $x_{M+1}=x_M+ \De$.
As for the mapping in \eqref{eq:xhat}, the mapping in \eqref{eq:xhat 2} is a one-to-one  correspondence between $W_k$ and $\Xh^{(k)}$.
Finally, let $\Sh^{(k)}=\Xh^{(k)}-U$ and $\Svh=[\Sh^{(k)}, \ldots, \Sh^{(K)}]$.
%
%
%The message estimate for the quantized message input is then obtained as
%\ea{
%\Xh=  \f 1 {|\hv^{(K)}|_2^2 }\sum_{k=1}^K h_k^2 \Xh_k-U.
%}
%
%
%\ean{
%\Sh^{(k)}& =\f 1 {|\hv^{(K)}|_2 }\sum h_k^2 \Yh_k-U.
%%& =h_k S + Z_k+h_k N_k
%}
%
%for $\Sh^{(k)}$  defined in an analogous manner as $\Xh$ in \eqref{eq:xhat} so that the message estimate is produced as
%\ean{
%\Xh
%& =\f 1 {|\hv^{(K)}|_2 }\sum h_k \Yh_k,
%}
%
%\blue{LUCA: Shouldn't we first build
%\ean{
%	\Xh
%	& =\f 1 {|\hv^{(K)}|_2^2 }\sum h_k \Yh_k,
%}	
%and then get $\widehat{S} = \widehat{X}-U$?
%	}

\medskip
%\noindent
%{\bf Analysis:}
We next lower bound the achievable rate as follows: first (i) we show that the capacity of the channel with finite quantization levels is to within a
constant gap from the channel with infinite quantization levels, successively (ii) we lower bound  that the capacity of the model with infinite quantization levels.
This lower bound minus with the gap between the capacity of the model with finite and infinite quantization corresponds to the achievable rate in \eqref{eq:achievable simo multiple 1}.

Define $\Xt^{(k)}$ as the quantization of  $W_k$ for $k \in [1, \ldots ,K]$ with infinite quantization levels and with step $\De$ as in \eqref{eq:quantized gaussian}.
Similarly, let $\St^{(k)}=\Xt^{(k)}-U$ and $\Svt=[\St^{(1)}\ldots \St^{(K)}]$.

The rate 	achievable with the transmission strategy described above is lower bounded as
%
%can be lover bounded as
\eas{
R^{\rm IN}% &= I(\Yv;X) \\
%&= I(Y_1^K;X|U) \\
& \geq  I(\Xvh; X) \\
%& \geq  I(\Xvh; S) \\
& =  H(\Xvh,\Xvt)-H(\Xvt|\Xvh)-H(\Xvh|X)\\
%&  \geq  H(\Svh,\Svt)-H(\Svt|\Svh)-H(\Svt|S)\\
& = I(\Xvt; X)-H(\Xvt|\Xvh) \\
& \geq  I(\Xvt; X)-\sum_{k=1}^K H(\Xt^{(k)}|\Xh^{(k)}).
\label{eq:gap mi}
%& \geq  I(\Svh,\Svt; S) \\
}
The expression in \eqref{eq:gap mi} is interpreted as follows: $I(\Xvt; X)$ is the attainable rate for the model with infinite output quantization while
$\sum_{k=1}^K H(\Xt^{(k)}|\Xh^{(k)})$ is an upper bound to the performance gap between the attainable rate with infinite and finite quantization.

\medskip

Let us first bound the performance gap between the channel with finite and infinite output quantization:
for each term $H(\Xt^{(k)}|\Xh^{(k)})$, we observe that, if $W_k/h_k \in [x_1-\De/2,x_M+\De/2]$, then  $\Xt^{(k)}=\Xh^{(k)}$:
using this observation and given the symmetry of the input and noise distributions, we write
\ea{
-H(\Xt^{(k)}|\Xh^{(k)})
& = -H(\Xt^{(k)}|\Xh^{(k)}=x_{M+1})  \Pr[\Xh^{(k)}_i=x_{M+1}] -H(\Xt^{(k)}|\Xh^{(k)}=x_0)  \Pr[\Xh^{(k)}=x_0] \nonumber \\
& = -2 H(\Xt^{(k)}|\Xh^{(k)}=x_{M+1})  \Pr[\Xh^{(k)}_i=x_{M+1}].
\label{eq:bound entropy single}
}
If $i \in [0,M+1]$, then
\ea{
\Pr[\Xt^{(k)}=x_i|\Xh^{(k)}=x_{M+1}]=0,
}
on the other hand, for $i >M+1$, we have
\ean{
\Pr[\Xt^{(k)}=x_i|\Xh^{(k)}=x_{M+1}]
%& \leq \max \lcb \Pr[\Xt^{(k)}=k|X_i=0], P[\Xt^{(k)}=k|X_i=x_{M+1}]\rcb  \\
& =\sum_{m=1}^M \Pr[\Xt^{(k)}=x_i|\Xh^{(k)}=x_{M+1}, X=x_m] \Pr[X=x_m] \\
&  \leq \f 1 M \sum_{m=1}^M \Pr[\	Xt^{(k)}=x_i| \Xh^{(k)}=x_{M+1}, X=x_M] \\
& \leq  \Pr[\Xt^{(k)}=x_i|\Xh^{(k)}=x_{M+1},X=x_{M}] \\
& \leq \f {Q \lb h_k((i-1)\De+\De/2)\rb-Q \lb h_k(i\De+\De/2))\rb}{ {Q( h_k M \De+\De/2)} } \\
%%
%& \leq  Q(h_i((k-1)\De+\De/2))\\
%%
& \leq  \f {Q \lb h_k((i-1)\De+\De/2\rb}{Q( h_k M \De+\De/2)},
\label{eq:last ineq} \\
& \leq \lb 1 -\f 1 {h_k (M \De+\De/2)} \rb^{-1} e^{- h_k^2(i^2-M^2)\De^2}  \\
& \leq \lb 1 -\f 1 {h_k M \De} \rb^{-1} e^{- h_k^2(i-M)^2\De^2}
}	
The case for $i<0$, can be bounded in a symmetric matter to yield
\ea{
\Pr[\Xt^{(k)}=x_i|\Xh^{(k)}=x_{M+1}] \leq
 \f {Q \lb h_k((i-1)\De+\De/2\rb}{Q( h_k M \De+\De/2)} \leq   \lb 1 -\f 1 {h_k M \De} \rb^{-1} e^{- h_k^2(i-M)^2\De^2}.
%  Q \lb h_k(1-i)\De \rb, \quad i<0
}
Since $M^2 \De^2=12 P$ and $P h_k>1$ by assumption, we have
\ea{
h_k M \De \geq   2 \sqrt{3}, \quad k\in[1, \ldots, K],
}
which implies $ Q((i h_k \De)) \leq e^{-1}$ for all $k$.
%\ea{
% Q((i h_k \De)) \leq e^{-1}, \quad k \in \Zbb \setminus \{1...M\},
%}
Since $-x \log x$ is positive increasing function in $x$ for $x \in [0,1/e]$, we can write
\eas{
& H(\Xt^{(k)}=x_m|\Xh^{(k)}=x_{M+1}) \\
& =  \sum_{i=M}^\infty  \Pr[\Xt^{(k)}=x_i|\Xh^{(k)}=x_{M+1}]  \log  \Pr[\Xt^{(k)}=x_i|\Xh^{(k)}=x_{M+1}] \nonumber \\
& \quad \quad + \sum_{i=0}^{-\infty}  \Pr[\Xt^{(k)}=x_i|\Xh^{(k)}=x_{M+1}] \Pr[\Xt^{(k)}=x_i|\Xh^{(k)}=x_{M+1}] \log  \\
& \leq  2 \sum_{i=M}^\infty  Q(k h_i \De)  \log  Q(k  h_i \De)   \\
%
%& \geq (1/4)*(2*exp(-(1/2)*M^2*De^2)*M*De-sqrt(Pi)*sqrt(2)*erf((1/2)*sqrt(2)*De*M)+sqrt(Pi)*sqrt(2))/(De*ln(2))
& \leq  \sum_{k=M}^\infty  \sum_{i=M}^\infty  \Pr[\Xt^{(k)}=x_i|\Xh^{(k)}=x_{M+1}]  \log  \Pr[\Xt^{(k)}=x_i|\Xh^{(k)}=x_{M+1}] \\
%%
%& \geq - \f 1 {2\log(2)} M e^{-M^2 h_i^2 \De^2} \\
%%
%\nonumber \\
%\nonumber \\
& \leq  \sum_{j=0}^\infty    \f 6 5 h_k^2 j^2\De^2 e^{- h_k^2 j^2\De^2}
\label{eq:exp bound}
\leq 0.15,
%& \geq - \f 1 {2\log(2)} M e^{- \f M 2  }
}{\label{eq:bound entropy single given}}
where we have used the fact that
\ea{
	\sqrt{15 P} \leq M \De \leq \sqrt{18 P},
}
and, similarly, $\sqrt{12} \leq \De \leq \sqrt{15}$ for $P>6$.

Plugging the bound in \eqref{eq:exp bound}  in \eqref{eq:bound entropy single} yields
\ean{
\sum_k H(\Xt^{(k)}|\Xh^{(k)})  \leq 0.3,
}
which shows that the capacity of the channel with infinite quantization is at most $0.3 \ \bpcu$ from the capacity of the channel with finite output quantization.
%%
%\newpage
%
%
%Note that this result hinges on the fact that two extra quantization levels are used to identify the event in which the channel output is below the minimum and maximum range of the
%channel input.

\medskip
Having bounded the performance gap between finite infinite quantization, we next lower bound the rate attainable in the model with infinite quantization of the $K$ antenna outputs with the highest SNR.
For this model, the attainable rate can be lower bounded using the results that Gaussian distributed noise is the worst additive noise under a covariance
constraint in \cite{diggavi2001worst}.
More specifically, let us define
%From classic properties of dithered quantization \cite{zamir1996lattice}, we have
%\ea{
%\Yh_k^{\infty}& =h_k S + Z_k+h_k N_k,
%\label{eq:channel inf}
%}
\ea{
\St& =\f 1 {|\hv^{(K)}|_2^2 }\sum h_k^2 \St^{k}-U.
}
Note that, from properties of dithered quantization \cite{zamir1996lattice}, we have
\ea{
\St^{(k)}& =S + \f {Z_k} {h_k}+ N_k,
}
where $N_k\sim \Ucal( \De/2, \De/2)$ and independent from $S$ and $Z_k$.
Using this observation, we have
\ean{
I(\Xt;X) & = I(\Xt;X|U) \\
&  \geq  I(\St;S) \\
& \geq  I(S+Z^{\Ncal};S),
}
where $Z^{\Ncal}\sim \Ncal(0,\ga)$ for
\ea{
\ga=\f 1 {|\hv^{(K)}|_2^4 } \lb \var \lsb \sum_{k=1}^{K} h_k  Z_k \rsb+ \var \lsb \sum_{k=1}^{K} h_k^2  N_k \rsb \rb
\label{eq:gamma}
}
Note that, from the achievability proof in Prop. in \ref{prop:multilevel SISO}, we  have
\ea{
 I(S+Z^{\Ncal};S) \geq \log(M)-0.6-\log(\ga).
 \label{eq:inner bound M }
}
A bound on $\ga$ in \eqref{eq:gamma} is obtained as follows:
\ea{
\var\lsb \sum_{k=1}^K h_k Z_k  \rsb=|\hv^{(K)}|_2^2
\label{eq:var 1}
}
and
\ea{
\var\lsb \sum_k h_k^2 N_k  \rsb \leq \f 1 {12} |\hv^{(K)}|_4 + \f 2 {12} \prod_{i>j} h_i^2 h_j^2 \leq \f {|\hv^{(K)}|_4^2} {12}
%\lb \sum_k h_k \rb^2 \De^2 \leq  \f {|\hv^{(K)}|_1^2} {|\hv^{(K)}|_2^2}
%|h|_1^2 \f {\De^2} {12}  \leq \f 1 {12 } \f {12 P}  {M^2-1}
\label{eq:var 2}
}
so that,  $h_i>1$, as by assumption
\eas{
\ga & \leq  \f {|\hv^{(K)}|_2^2+|\hv^{(K)}|_4^2} {|\hv^{(K)}|_2^4}  \\
& \leq 1+\lb \f{|\hv^{(K)}|_4} {|\hv^{(K)}|_2^2}\rb^2 \leq 2.
}{\label{eq:bound gamma}}

Substituting $M$ in \eqref{eq:choose M 2} and bounding $\ga$ as in \eqref{eq:bound gamma} in
\eqref{eq:inner bound M } finally yields \eqref{eq:achievable simo multiple 1}.

\section{Proof of Prop. \ref{prop:mimo single}}
\label{app:mimo single}
With single antenna selection, the capacity maximization in \eqref{eq:max c} can be rewritten as
\ea{
\Ccal(\Fcal) \leq \max_k \f 12 \log \lb 1+|\hv_k|_2^2 P, (N_{SQ}+1)^2\rb,
\label{eq:max sas}
}
where $\hv_k$ is the $k^{\rm th}$ row of $\Hv$.
In other words, the capacity is the maximum among the capacity of the MISO channels between the transmitter and each of the antennas at the receiver.
For each MISO channel, the capacity can be attained using the result in Prop. \ref{prop:simo multi} since transmitter pre-coding can be used to turn the MISO channel into a SISO channel.

%
%
%\newpage
%
%
%%since the capacity of each antenna choice can be bounded using Prop. \ref{prop:multilevel SISO}.
%%
%The function on the RHS of \eqref{eq:max sas} is increasing in $k$ so that \eqref{eq:mimo_b} is obtained. \blue{LUCA: Don't we have an achievable scheme? In Prop. 6 we claim that the gap from capacity is at most 2 \bpcu}

\section{Proof of Prop. \ref{prop:mimo linear}}
\label{app:mimo linear}
%as in $L$
%and the overall rate
Through the classic VBLAST architecture, the channel can be equivalently written as a set of parallel channels
\ea{
\Wt_i=\la_i \Xt_i+\Zt_i, \quad i=1,\ldots, \min\{\Ntx,\Nrx \},
\label{eq:SVD channel}
}
where $[\la_1, \ldots, \la_{\min\{\Ntx,\Nrx \}}]$ are the eigenvalues of $\Hv$ and $\Zt_i \sim \iid  \Ncal(0,1)$.

Since the capacity of the parallel of  channels is obtained as the sum of the capacity of each channel, we have that an upper bound to capacity is
\ea{
R^{\rm OUT}=\max  \sum_{i=1}^{\min\{\Ntx,\Nrx  \}} \f 12 \log \lb \min \{ \la_i^2 P_i+1,(N_{SQ,i}+1)^2\} \rb,
\label{eq:mimo max}
}
where the maximization is over $\sum P_i=P$ and $\sum N_{SQ,i}=N_{SQ}$ as $P_i$ is the input power  and $N_{SQ,i}$ the number of sign quantizers allocated to the $i^{\rm th}$ equivalent channel.
Additionally, the upper bound in \eqref{eq:mimo max} can be attained to within $\min\{\Ntx,\Nrx \} \ \bpcu$ following the result in Prop.~\ref{prop:multilevel SISO}.
%
%\blue{LUCA: Update this sentence if the new gap from capacity of SISO is larger than 0.5 \bpcu}
% DONE
%Note that necessarily capacity  \blue{LUCA: sentence missing}

We next wish to determine an approximate expression for the solution of the optimization in  \eqref{eq:mimo max} as a function of the available power and number of sign quantizers.
To simplify this analysis, we relax this optimization problem and let $N_{SQ}$ take values in $\Rbb^+$.
Under this relaxation of the optimization problem in \eqref{eq:mimo max}, we have that the term  $\min \{ \la_i^2 P_i+1,(N_{SQ,i}+1)^2\}$
must be attained by either the power or the sign quantizer allocation on all channels simultaneously.
This can be shown by contradiction: assume that there exist two subchannels $j$ and $k$ such that
\eas{
\min \{ \la_j^2 P_j^*+1,(N_{SQ,j}^*+1)^2\} & = \la_j^2 P_j^*+1 \\
\min \{ \la_k^2 P_k^*+1,(N_{SQ,k}^*+1)^2\} & = (N_{SQ,k}^*+1)^2,
}
in the optimal solution, then there must exist $\ep_1,\ep_2>0$ such that
\eas{
\min \{ \la_j^2 (P_j^*+\ep_2)+1,(N_{SQ,j}^*-\ep_1+1)^2
\} & = \la_j^2 (P_j^*+\ep_2)+1>  \la_j^2 P_j^*+1\\
\min \{ \la_k^2 (P_k^*-\ep_2)+1,(N_{SQ,k}^*+\ep_1+1)^2\} & = (N_{SQ,k}^*+\ep_1+1)^2> (N_{SQ,k}^*+1)^2,
}
which contradicts the claim of optimality.
For the case in which the power constraint is active, the optimal solution corresponds to the classical waterfilling solution
in the channel with infinite quantization levels.
For the case in which the constraint on the quantization is active, then maximization problem becomes
\ea{
\max_{\sum N_{SQ,i}=N_{SQ}}  \sum_{i=1}^{\min\{\Ntx,\Nrx  \}}\log \lb N_{SQ,i}+1 \rb.
\label{eq:max 2}
}
The optimization problem in \eqref{eq:max 2} is equivalent to the waterfilling problem with equal channel gains and thus
 the uniform allocation of quantizers across all sub-channels is optimal.
%
%\blue{LUCA: I don't see why}, in which case
% SR: should be ok now!
%
\ea{
1+\la_i^2 P_i=\lb \f {N_{SQ}}{K}+1 \rb^2
%\la_i P_i+1=\lb \rb
}
where $K$ is the number of active channels.
%
%Since
%\ea{
%1+\la_i P_i \geq \lb \f {N_{SQ}}{K}+1 \rb^2.
%}
Since $N_{SQ}+1\leq 2 N_{SQ}$, the assignment $\lfloor N_{SQ} \rfloor$ provides a loss of at most $1 \ \bpcu$ per each channel, so that the overall gap between inner and upper bound is $2 \min\{\Ntx,\Nrx \} \ \bpcu$.
\end{document}